\documentclass[preprint]{elsarticle}

\usepackage{graphicx,subfigure}
\usepackage{epstopdf}
\usepackage{amsmath, amssymb}
\usepackage{rotating}

\begin{document}
\title{Momentum universe shrinkage effect in price momentum}
\author[sb]{Jaehyung Choi\corref{cor1}}
\ead{jaehyung.choi@sunysb.edu}

\author[qnq]{Sungsoo Choi}
\ead{schoi@qnqgroup.com}

\author[qnq]{Wonseok Kang}
\ead{wonseok.kang@qnqgroup.com}

\cortext[cor1]{Correspondence address: Department of Physics and Astronomy, SUNY at Stony Brook, NY 11794-3800, USA. Fax:+1-631-632-8176.}
\address[sb]{Department of Physics and Astronomy, SUNY Stony Brook, NY 11794-3800, USA}
\address[qnq]{Q\&Q Investment Group, Seochodong Seochogu, Seoul, 137-772, Korea}

\begin{abstract}
	We test the price momentum effect in the Korean stock markets under the momentum universe shrinkage to subuniverses of the KOSPI 200. Performance of the momentum strategy is not homogeneous with respect to change of the momentum universe. It is found that some submarkets generate the higher momentum returns than other universes do but large-size companies such as the KOSPI 50 components hinder the performance of the momentum strategy. The observation is also cross-checked with size portfolios and liquidity portfolios. Transactions by investor groups, in particular, the trading patterns by foreign investors can be a source of the momentum universe shrinkage effect in the momentum returns.
\end{abstract}
\begin{keyword}
	momentum strategy, ranking criterion, momentum universe shrinkage
\end{keyword}
\maketitle
\section{Introduction}
	Since discovery of the momentum and reversal effects in stock markets of various time scales \cite{Jegadeesh:1993p200, Asness:1994, DeBondt:1985p4232, Lo:1990p883}, the momentum effect has become one of the most controversial issues in finance for testing robustness of the efficient market hypothesis, originally suggested by Fama \cite{Fama:1965p4897} and Samuelson \cite{Samuelson:1965p4904}. As a cornerstone principle in finance, the efficient market hypothesis claims that any kinds of systematic trading strategies cannot outperform their benchmarks, such as market indexes, in the long run. However, the momentum strategies in financial markets of numerous asset classes \cite{Rouwenhosrt:1998,Rouwenhosrt:1999,Okunev:2003,Erb:2006,Asness:2008,Moskowitz:2010} are the counter-examples to the hypothesis not only because simply looking at past winners and losers can forecast future directions of price changes but also because the expected risk-adjusted returns of the momentum strategies in various asset markets are statistically significant non-zero values indicating that the strategies beat market benchmark indexes in which the strategies are implemented.
	
	The more exotic fact on the momentum effect is that the factors driving the price momentum in financial assets don't have been clearly revealed yet. The failure of the Fama-French three factor model to explaining the momentum returns \cite{Fama:1996} has opened various questions to origins of the momentum effect. The sector momentum plays a role of a partial answer to the momentum source \cite{Moskowitz:1999p4294}. The behavioral approaches, such as investors' behaviors including under/over-reactions of investors to news and investors' sentiment, provide another edge of the answer \cite{Hong:1999p4506,Terence:1998p4385,Daniel:1998p4514,Barberis:1998p307} and those attempts now become another research stream in finance, also known as behavioral finance. Cross-correlation \cite{Lo:1990p883} and autocorrelation \cite{Lewellen:2002} explain some portion of the momentum effect. Meanwhile, it is claimed that the momentum effect is an illusory phenomenon by transaction costs \cite{Lesmond:2004}. However, there is not the complete answer to the unified framework for explaining the reason why we have had the momentum effect over a few decades in various asset markets.

	As mentioned above, many researchers have tried to unravel the origin of the momentum effect because it is able to directly answer to robustness of the efficient market hypothesis. On the other hand, seeking the characteristics and origins of the momentum returns is also practically helpful to making money in real markets. In particular, introduction of new ranking criteria can provide more profitable momentum strategies. Asness et al. \cite{Asness:2008} suggested that there are negative correlations between value and momentum across various asset classes and a marriage of the momentum strategy with value investment can increase profitability of the momentum strategy. As an instance, it is impossible to find any statistically meaningful signal for the momentum returns in the Japanese stock market \cite{Asness:2011}. However, the marriage with value investment, that calculates the intrinsic values of firms by using book-to-market ratio and other metrics on accounting, ameliorates performance of the momentum strategy \cite{Asness:2008}. If more criteria from the value investment and/or other valuation parameters from totally different origins are plugged into the momentum model, it is probable to upgrade the performance of the momentum strategy and to find some clues to the origin of the momentum effect which tests robustness of the efficient market hypothesis.
	
	Asness' work is not the only approach to introducing new ranking criteria for the momentum portfolios. Rachev et al. \cite{Rachev:2007p616} used various risk measures such as Value-at-Risk, Sharpe ratio, STARR ratio, and R-ratio in order to select stocks for the momentum style portfolio. They also suggested that each of momentum groups has its own risk properties and the risk groups encrypt the dependencies on momentum effects. In their work, the momentum strategies based on the risk measure criteria improve the momentum returns and have lower tail risks than the cumulative return based momentum. Additionally, Lee and Swaminathan \cite{Lee:2000} considered the volume effect on the momentum phenomenon. They found that low past volume stocks tend to have the higher future returns but the momentum strategies become more profitable with large volume equities. Moreover, George and Hwang \cite{George:2004} and Liu et al. \cite{Liu:2011} employed the 52-week high price as a momentum ranking rule which can generate the better momentum returns than the traditional momentum strategies in the U.S. and various international markets, respectively. The momentum strategy with time-series forecasting on the return is also proposed by Moskowitz et al. \cite{Moskowitz:2010} and the momentum returns are improved by their time-series model. The analogy of price momentum is also suggested from physics in order to find more profitable momentum strategies \cite{Choi:2012a}.
	
	Although most of the literatures with various approaches have tested the existence of the momentum effect in diverse market universes and have figured out the origin and profitability of the momentum strategies, none of attentions to their subuniverses for analysis are not given at all. For examples, although many momentum papers cover all enlisted equities in local stock exchanges, it is not narrowed down to the Russell 3000, or the S\&P 500 as major market universes nor their subuniverses such as the Russell 1000, the Russell 2000, or the S\&P 100, respectively. Since the subuniverses are not constructed by a random selection but chosen by construction rules, the construction rules for the subuniverses are important to give hints on the source of the momentum effect if the returns are influenced by the universe change. Some studies \cite{Hong:1999p4506,Yu:2008} took care of its subsets not for finding profitability of the momentum strategy in the subuniverses but for testing how future returns of equities are affected by investors' interests.
	
	In this paper, we test the momentum effect in the KOSPI 200 which is a subset of all enlisted equities in the South Korean stock market. After finding existence of the momentum effect in the KOSPI 200, the test is extended to its subuniverses which have different criteria for universe construction. By using these different construction rules, the indirect search for the momentum driving factors is also performed. The methodology is that the traditional momentum-style strategies with selection criteria such as market capitalization, liquidity, and transaction by investor groups are implemented. The results of these portfolios suggest that there are diverse characteristics which drive the price momentum in each subuniverse. This paper focuses on reporting a new phenomenon in the momentum effect. Instead of carrying out factor analysis at certain lookback-holding pairs or building up  behavioral models, the empirical study over various lookback-holding pairs is conducted in order to find the general tendency of the portfolio performance under the momentum universe shrinkage. 
	
	The structure of this paper is the following. In next section, we briefly cover dataset used in the study and the way how to implement the momentum strategy. The construction rules for each subuniverse and a basic glossary are also introduced. In section 3, the momentum returns for the subuniverses are compared with that for the original market pool, the KOSPI 200. From the comparison, the momentum universe shrinkage effect is found. In section 4, market capitalization based portfolios are implemented for explaining the results in section 3. In section 5, the liquidity is tested as a momentum factor and the shrinkage effect is observed with the liquidity portfolio. In section 6, strategies by transaction style of investors are back-tested over the real data and the transaction information can be used as proxies for explaining part of the market shrinkage in the momentum effect. In section 7, we close the paper.

\section{Dataset and methodology}
	The main market universe of our analysis is the KOSPI 200 in the South Korean stock market managed by the Korea Exchange (KRX). Since the dataset here is used by previous studies \cite{Choi:2012a, Choi:2011} which provide detailed descriptions on the dataset, a brief introduction is given here and consult with \cite{Choi:2012a, Choi:2011} for further information on the dataset. The KOSPI 200 is the main index that includes 200 sector-representative equities in the South Korean market. It is apparently thought as the main benchmark to the investors with the following reasons. First of all, it is the only market index which has options and futures on itself and it is also known that the Korean derivative market is one of the most liquid markets in global capital markets. Secondly, there exist numerous transactions on the KOSPI 200 index and its components while the market opens not only because they are correlated to the derivative markets but also because the KOSPI 200 contains qualified equities which are in good condition from the viewpoints of business performance and fiscal stability.
	
	The KOSPI 200 has two formal subsets of which the construction criteria are market capitalization and sector diversification. One of the explicit subsets is the KOSPI 100 that is a value-weighted index of the sector-diversified top 100 companies in size which are also components of the KOSPI 200. Another official subset is the KOSPI 50 which considers only the market capitalization as a construction rule for selecting the top 50 companies in the order of market capitalization. Similar to the KOSPI 200, both subsets are maintained and announced by the KRX prior to component replacement and rebalancing. With the three universes, it is possible to construct other implicit subuniverses by combination of these three universes. If the three universes are called F for the KOSPI 50, simply $(50)$, H  for the KOSPI 100 or $(100)$, and T for the KOSPI 200 or $(200)$, possible combinations for the complementary subsets are $T\cap H^c$ $(200-100)$, $T\cap F^c$ $(200-50)$, $H\cap F^c$ $(100-50)$, and $(T\cap H^c)\cup F$ $(200-100+50)$. All these seven universes are used for implementation of the momentum strategy.

	In addition to the criteria for the KOSPI 50 and the KOSPI 100, each complementary subuniverse has a meaning of construction. Usually, the construction rule weighs more on the market capitalization. For examples, components of $(200-50)$ are all middle-cap or small-cap companies and those of $(200-100)$ are considered as all small-sized equities. Moreover, $(100-50)$ contains purely middle-size firms. Meanwhile, the small-sized and large-cap companies not including the middle-cap equities are covered by $(200-100+50)$. Although the sector diversification is one of the universe-building rules for the KOSPI 100 and the KOSPI 200, its contribution to the complementary subsets is rather ambiguous because the KOSPI 50 is not dependent with sectors. As an instance, there is no guarantee that $(100-50)$ is sector-diversified since some sectors such as the telecommunication sector, which are covered only by small numbers of large-cap companies, are not in $(100-50)$.
	
	The time span of the dataset is from January 2000 to December 2011. During the period, following data are collected from the KRX: daily price data of the KOSPI 200 components, market capitalizations of these equities, daily transaction data of investor groups, and change logs of the KOSPI 200, the KOSPI 100, and the KOSPI 50 components. The price data are adjusted under consideration on corporate events such as stock split/reverse split, capital increase/decrease, and dividend payment. In the case of transaction data for each investor group, net amounts in volume and cash are collected. The change logs for the complementary subsets are simply obtained from combinations of the logs of the three main indexes. 
	
	The basic methodology is back-testing a momentum strategy in a given market universe and then verifying its performance. Since the details how to implement the momentum strategy is given in Jegadeesh and Titman \cite{Jegadeesh:1993p200}, it is explained shortly here. In a given universe, the momentum portfolio is constructed based on rankings by cumulative return during estimation periods. At $t=0$, the equities in the universe are looked back from $t=-J$ to $t=-1$ and sorted by the criterion in ascending order. After then, they are separated into ten groups and equities with the lowest rankings are called the losers which are in R1. Stocks which have the highest rankings are grouped to the winners or R10. With these two groups, we can buy the best (R10) and short sell the worst (R1) with the same amount to create the dollar-neutral portfolio. Weights of the equities in each winner and loser group are identical. After $K$ periods, the portfolio is liquidated to take profits from the momentum portfolio, by selling the R10 and buying the R1 back.
	
	For a basic glossary, the profitability means whether an expected raw return of the portfolio is positive or negative. It only matters a sign of the raw return. If the expected raw return of the portfolio is given by $r$, the profitability is defined by
	\begin{eqnarray}
		PF=\frac{r}{|r|}=\pm 1.\nonumber
	\end{eqnarray} 
	If the profitability is 1, the momentum strategy is implemented. Otherwise, the contrarian strategy, exactly opposite position to the momentum portfolio, is executed. However, in any cases, the actual portfolio returns in practice are not guaranteed by non-zero raw returns of the momentum portfolio because of a transaction cost. In the South Korean market, the transaction cost is about 35 bps for each of winner and loser baskets. For the momentum portfolio, 70 bps of the transaction cost are subtracted from the raw returns. Even though positive raw returns are expected, the actual returns from execution of the strategies could be negative. For considering this situation, an absolute (or implemented) return with the transaction cost needs to be calculated in order to verify whether or not implementation of a certain strategy is able to generate the profits in the real market. The implemented (or absolute) return is defined by 
	\begin{eqnarray}
		r_{I}=|r|-c\nonumber
	\end{eqnarray}
	where $c$ is the transaction cost. If the absolute return with the transaction cost is expected to be positive, the strategy, whether it is trend-following or contrarian, makes money. Consideration on the transaction cost is also crucial for testing the claim by Lesmond et al. \cite{Lesmond:2004} that the momentum effect is caused by the transaction cost.  From now on, the return means the absolute or implemented return with the cost unless it is called the raw return. The implemented Sharpe ratio with transaction costs is also calculated by using the absolute return and it is given by
	\begin{eqnarray}
		SR=\frac{r_{I}}{\sigma}\nonumber
	\end{eqnarray}
	where $\sigma$ is the standard deviation of the raw returns. Similar to the return case, the Sharpe ratio in this paper means the implemented Sharpe ratio with transaction costs.
	
\section{Momentum return and market universe}
	The monthly momentum strategy in the South Korean market has been investigated and it is well-known that there is no statistically significant momentum effect with the market universe which covers all enlisted equities in the market \cite{Liu:2011,Koh:1997,Ahn:2004,Chae:2009}. Meanwhile, it is also reported that statistically meaningful momentum returns are obtained \cite{Choi:2012a}. Comparing with the previous studies, main differences in Choi's work \cite{Choi:2012a} are followings. First of all, it only considers the KOSPI 200 components as a momentum universe while other works covers all enlisted equities in the Korean markets. Moreover, this fact imposes that a change of the momentum universe brings the more profitable momentum strategy. Secondly, some equities, which don't have enough numbers of trading days in estimation periods for calculation of ranking metrics, are temporarily neglected from construction of the portfolio although those are included again when those equities have enough trading dates during lookback periods. However, the effect of removal seems not too significant since numbers of the equities of temporal removal are less than 2\% of the KOSPI 200 components. To verify the effect of the momentum universe change and to be consistent with the literatures, the momentum strategies with all equities in the KOSPI 200 need to be tested. This is the starting point of this paper.
	
	\begin{figure}[!t]
		\subfigure[Profitability]{\includegraphics[width=6cm]{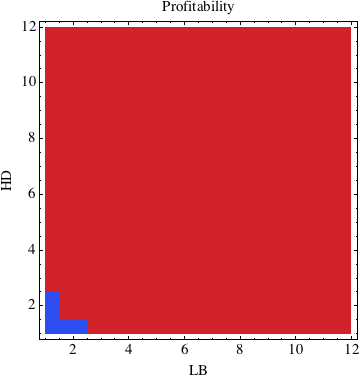}}
		\subfigure[Return]{\includegraphics[width=6cm]{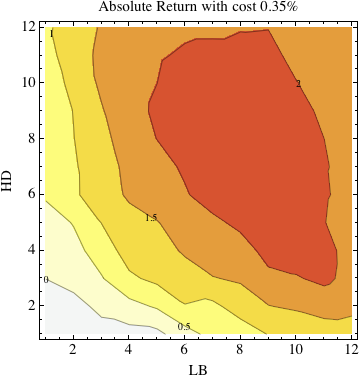}}
		\subfigure[Volatility]{\includegraphics[width=6cm]{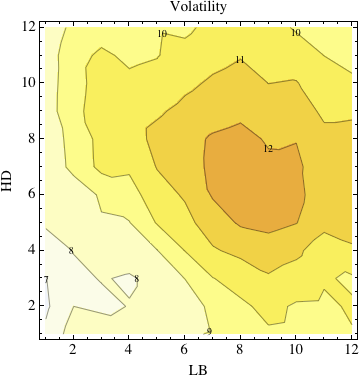}}
		\subfigure[Sharpe ratio]{\includegraphics[width=6cm]{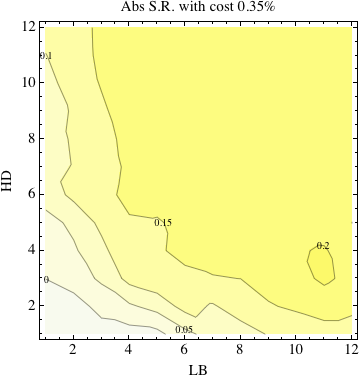}}
		\caption{Statistics for monthly momentum strategies in the KOSPI 200. All statistics are in the monthly scale.}
		\label{FIG_Monthly_Momentum_Stat}		
	\end{figure}

	The statistics of the monthly momentum strategies up to twelve months of estimation and holding periods are given in Fig. \ref{FIG_Monthly_Momentum_Stat}. As given in (a) of Fig. \ref{FIG_Monthly_Momentum_Stat}, most of the momentum strategies have positive average raw returns without transaction cost but short term strategies are reversal. This is well-matched to the original observation in Jegadeesh and Titman \cite{Jegadeesh:1993p200}. In the short terms corresponding to the weekly scale, it also shows the reversion of price which is consistent with Lo and MacKinlay \cite{Lo:1990p883}. In (b) of Fig. \ref{FIG_Monthly_Momentum_Stat}, the best strategy is located at the long term region and the strategies near the best strategy generate the plateau of more than 2\% even after subtracting the transaction cost of 35 bps per each of winner and loser baskets, totally 70 bps for the momentum portfolio. Although t-values are not in the paper, the null hypothesis, that the average return is zero, is rejected with confidence levels of 95\% or better for many strategies. One of the reasons why many long-term strategies outperform is that since the transaction cost is an one-time charge, the impact on monthly returnw is decreased as the holding period is stretched. The volatility graph, (c) in Fig. \ref{FIG_Monthly_Momentum_Stat}, shows that volatilities of the strategies are also increased as the holding period is lengthened. Although the Sharpe ratios in (d) are decreased by the increased volatility, the long-term strategies still have better Sharpe ratios. The monthly Sharpe ratios of many strategies are between $0.15-0.2$ corresponding to annually $0.5-0.7$.
	
	Comparing the previous studies on the momentum effect in the Korean markets, the results in Fig. \ref{FIG_Monthly_Momentum_Stat} impose that a change on the momentum universe actually gives an impact on performance of the momentum portfolio. The market universe used in the previous studies include the KOSPI 200 components in this paper. This shrinkage makes the momentum strategy more profitable and the returns are statistically significant even with the transaction cost. The momentum universe shrinkage onto the KOSPI 200 is not a random sampling from all enlisted equities in the market as described in the section on dataset. It is suggested that the construction rules of the KOSPI 200 are the candidates for the momentum factors. Additionally, it is already well-known that market capitalization and sector diversification are two famous momentum driving factors \cite{Fama:1996, Moskowitz:1999p4294}. We are intrigued by the shrinkage of the momentum universe to the subsets of the KOSPI 200 as the universe is narrowed down from all the enlisted equities in the KOSPI to the KOSPI 200.
	
	\begin{figure}[]
	\begin{center}
		\subfigure[(100)]{\includegraphics[width=3.6cm]{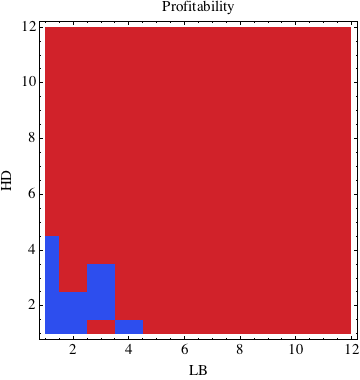}}
		\subfigure[(50)]{\includegraphics[width=3.6cm]{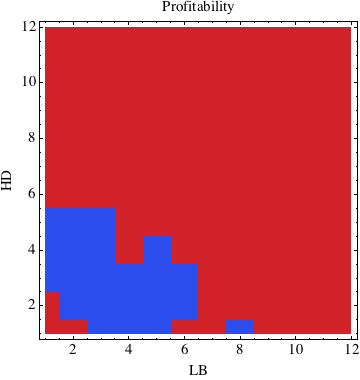}}
		\subfigure[$(200-50)$]{\includegraphics[width=3.6cm]{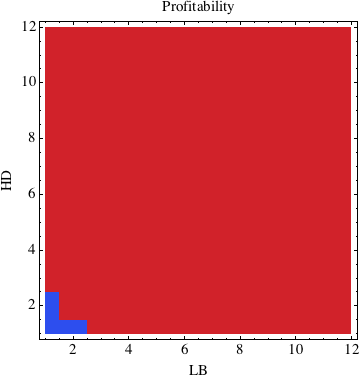}}
		\subfigure[$(200-100)$]{\includegraphics[width=3.6cm]{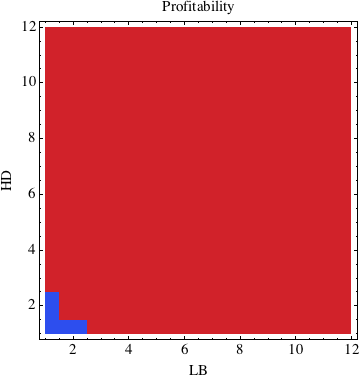}}	
		\subfigure[$(200-100+50)$]{\includegraphics[width=3.6cm]{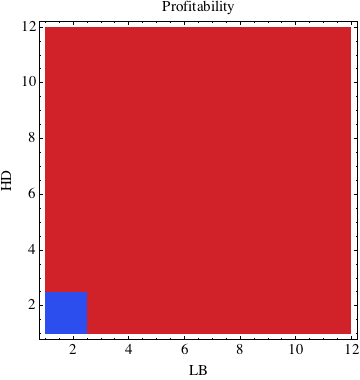}}		
		\subfigure[$(100-50)$]{\includegraphics[width=3.6cm]{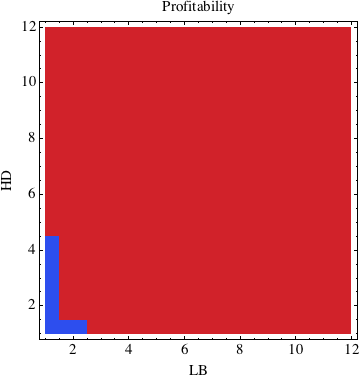}}		\end{center}
		\caption{Profitabilities of monthly momentum strategies in the subuniverses}
		\label{FIG_Subset_Profitability_M}		
	\end{figure}

\begin{figure}[]
		\subfigure[(100)]{\includegraphics[width=6cm]{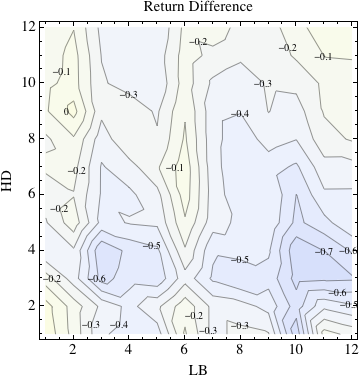}}
		\subfigure[(50)]{\includegraphics[width=6cm]{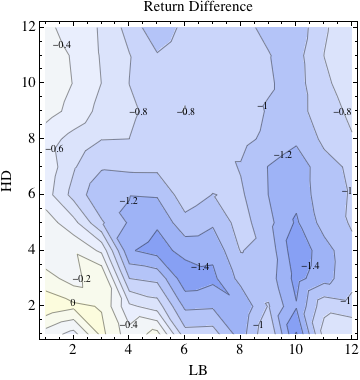}}
		\subfigure[$(200-50)$]{\includegraphics[width=6cm]{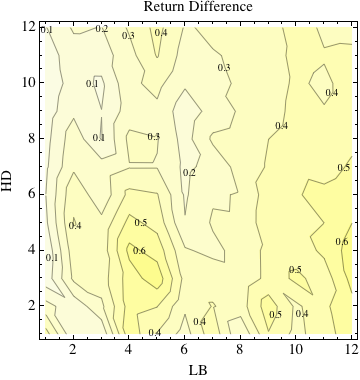}}
		\subfigure[$(200-100)$]{\includegraphics[width=6cm]{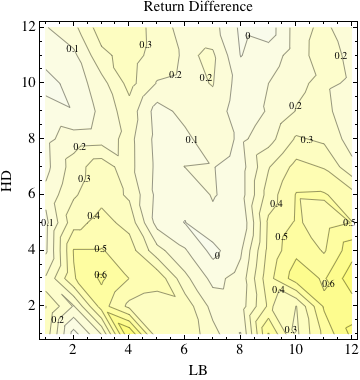}}	
		\subfigure[$(200-100+50)$]{\includegraphics[width=6cm]{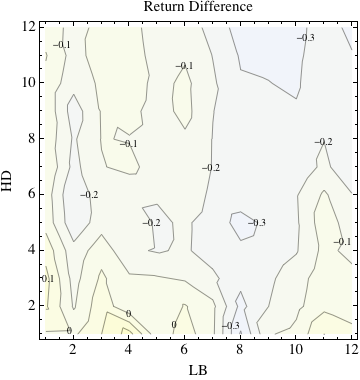}}		
		\subfigure[$(100-50)$]{\includegraphics[width=6cm]{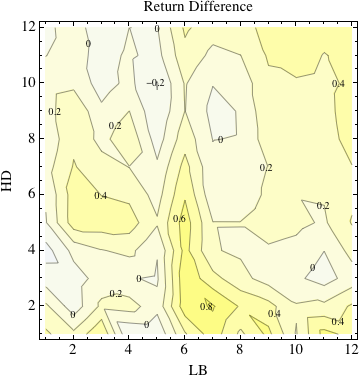}}		
		\caption{Relative returns of monthly momentum strategies in the subuniverses. All relative returns are in the monthly scale.}
		\label{FIG_Subset_Relative_Return}		
	\end{figure}

	The momentum strategies are repeatedly implemented over the subsets of the KOSPI 200. The results are interesting because the shrinkage provides not only more profitable strategies than the momentum strategy in the KOSPI 200 but also useful information on the origin of the momentum effect. Given in Fig. \ref{FIG_Subset_Profitability_M}, all subuniverses have trend-following strategies at almost all holding and lookback pairs except the short terms. The outcomes for relative strength to the KOSPI 200 momentum strategies, shown in Fig. \ref{FIG_Subset_Relative_Return}, are statistically significant. Most of the momentum strategies in the KOSPI 100 and the KOSPI 50, depicted in (a) and (b) respectively, are not as good as the momentum strategies in the original market pool, the KOSPI 200. Although the results of the KOSPI 100 are slightly better than those of the KOSPI 50, the momentum returns in the both markets are relatively weaker than those of the KOSPI 200 and other subsets. With consideration on the construction rules, these results propose that market capitalization and sector diversification are main factors for the momentum effect as it is already known \cite{Fama:1996,Moskowitz:1999p4294}.
 		
	\begin{figure}[]
		\subfigure[$(100)$]{\includegraphics[width=6cm]{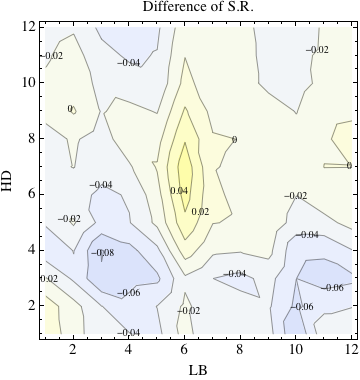}}
		\subfigure[$(50)$]{\includegraphics[width=6cm]{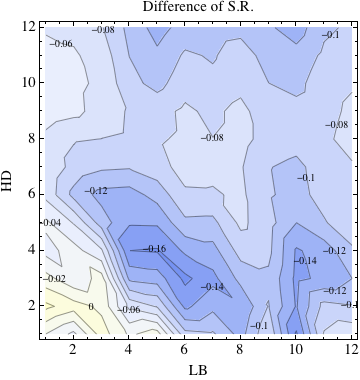}}
		\subfigure[$(200-50)$]{\includegraphics[width=6cm]{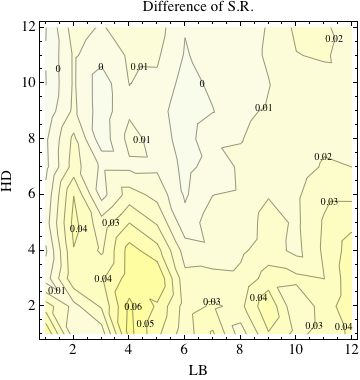}}
		\subfigure[$(200-100)$]{\includegraphics[width=6cm]{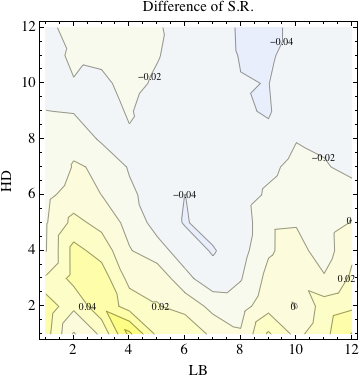}}		
		\subfigure[$(200-100+50)$]{\includegraphics[width=6cm]{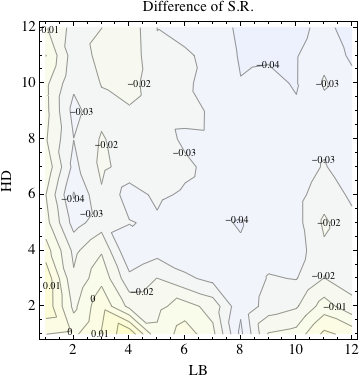}}		
		\subfigure[$(100-50)$]{\includegraphics[width=6cm]{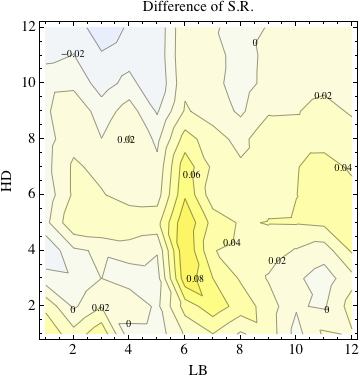}}				
		\caption{Relative Sharpe ratios of monthly momentum strategies in the subuniverses. All relative Sharpe ratios are in the monthly scale.}
		\label{FIG_Subset_Relative_SR}		
	\end{figure}
	
	The more interesting fact is that the KOSPI 50 gives a negative effect on the momentum performance. As far as the KOSPI 50 components are included into the momentum pools, the momentum strategies in those universes underperform the momentum strategies in other non-KOPSI 50 universes. For examples, $(200-50), (200-100),$ and $(100-50)$ beat the KOSPI 200 and the KOSPI 100. When the KOSPI 200 is compared with $(200-50)$ and $(200-100)$, the return differences of monthly 0.1\%$-$0.6\% are very outstanding because the best return in the KOSPI 200 is about monthly 2\%. Additionally, $(200-100+50)$, which also includes the KOSPI 50 components, has the weaker performance, less profits of monthly 0.3\%$-$0.6\%, than $(200)$ and $(200-100)$ although $(200-100)$ outperforms $(200)$. Meanwhile, a comparison between $(100-50)$ and $(100)$ tells that additional 0.1\%$-$1.0\% returns in a month are brought by exclusion of the KOSPI 50 components.
	
	The Sharpe ratios for each subuniverse, given in Fig. \ref{FIG_Subset_Relative_SR}, also supports the results in Fig. \ref{FIG_Subset_Relative_Return}. Similar to the conclusion from Fig. \ref{FIG_Subset_Relative_Return}, existence of the KOSPI 50 components in the momentum universe makes the strategies underperform the original momentum strategies in the KOSPI 200. The more interesting result is that $(200-50)$ clearly outperforms $(200-100)$ in Share ratio although it is hardly possible to discern the difference in relative returns between two universes. This fact imposes that the strategies in $(200-50)$ have smaller volatilities than the strategies in $(200-100)$ and also exhibit more consistent performance. Since the differences in composition between $(200-50)$ and $(200-100)$ are the components in $(100-50)$, it is guessed that equities in $(100-50)$ are useful for portfolio construction of better performance and Sharpe ratio than equities in $(50)$ are. The inclusion of the KOSPI 50 components is not good at decreasing volatility of the momentum portfolio.
	
	From the previous paragraphs, it is apparent that consideration on the KOSPI 50 components is not a good choice for increasing momentum profits and for decreasing volatilities of the portfolio. When the KOSPI 50 is included, the strategies underperform the KOSPI 200 momentum. However, if it is excluded, the momentum effect becomes more meaningful and more profitable. The dependence of the performance on existence of the KOSPI 50 components in the momentum universe looks significant and intrinsic. It is obvious that selection of the market universe has important meanings not only in improving the performance but also in searching for the origin of the momentum effect. Considering how the KOSPI 50, the KOSPI 100, and the KOSPI 200 are constructed, it is indirectly deduced that the momentum effect is largely based on sector momentum and market capitalization. Further analysis on this direction will be given in the next section.
	
\section{Market capitalization and size portfolio}
	In the previous section, the dependence of the momentum returns on market universe is found. Although the market universes are subsets of the KOSPI 200, the patterns in the momentum returns and Sharpe ratios over the lookback and holding pairs are divergent as the universes are altered. Additionally, some subsets such as the KOSPI 50 seriously influence performance of the momentum strategy. One of possible explanations on this observation can be from the rules of its subuniverse construction. Since each subuniverse has its own construction rules which consider one of market capitalization and sector diversification or both of those as explained before, more attention should be paid on these factors in order to increase the momentum returns and to clarify the source of the momentum effect. 
	
	However, the market capitalization is only considered in this paper. The reasons are followings. First of all, our main concern, the KOSPI 50, is not sector-diversified but is constructed only by the market capitalization. Unlike other subsets, the sector momentum seems not to be an important driving factor under the market shrinkage because the sector diversification is not a matter of concern in the KOSPI 50. Second, there is no concrete guarantee that some originally sector-diversified universes give still sector-diversified subsets after excluding the components of the KOSPI 50. Since sizes of the market universes are relatively smaller than the other literatures, it is probable that some sectors have really small numbers of companies or no companies in certain sectors. The example is the telecommunication sector in $(100-50)$. Third, the KOSPI 200 is already highly concentrated on several sectors because large portions of the KOSPI 200 components are in the information technology (IT) sector. More than half of components are IT-related because lots of companies are doing business in semi-conductor related industries. That is why the IT sector is more weighted than other sectors in the South Korean market.
	
	In particular, the market capitalization is one of the most important characteristics in the firm valuation. As one of the three factors in the Fama-French model \cite{Fama:1996}, its importance always has gotten the attention from the academics and the momentum effect is also partly dependent with the firm size. The large companies can give more serious impacts on the index and are able to change the overall sentiment of the markets and investors. Moreover, large market capitalization companies usually attract more interest of investors and this tendency is also found in Boni and Womack \cite{Boni:2006} that analysts tend to write more analysis reports which can disseminate more information on the larger firms. Meanwhile, the main index and market sentiment are weakly correlated with the price changes of the small-sized companies. Additionally, very small numbers of analysts only cover the small-cap companies.
	
	As mentioned in the section for dataset, one of the two main differences between its subuniverses is market capitalization. Among the KOSPI 200 components, the KOSPI 50 is constructed from the top 50 companies in market capitalization without consideration on sectors. Among the KOSPI 200 members, components of the KOSPI 100 are the equities in the upper-half class in market capitalization. The results from other subuniverses derived from the complementary sets are explained in terms of market capitalization. For examples, $(100-50)$ and $(200-100)/(200-50)$ contain the smaller- and/or middle-sized firms in the KOSPI 100 and the KOSPI 200, respectively.
	
	Considering the difference in market capitalization of each submarket, a long-short portfolio, based on market capitalization as a ranking criterion, is constructed. On the day of portfolio construction, all components in a given market pool are sorted by the size and then are classified into ten groups. The smallest companies are allocated into the group 1 (S1) and the largest companies into the last group (S10). After then, the long-short portfolio is constructed by buying the large caps (S10) and short-selling the small caps (S1). Any dependence of the momentum effect on market capitalization in a given market universe can be caught by the return of this market capitalization portfolio if it exists. When the expected return is positive, the large companies outperform the small companies. If it is negative, the small firms are preferred for generating revenues and it is well-matched to the prior observation in Kim et al. \cite{Kim:2012}.
	
	\begin{figure}[]
		\subfigure[monthly]{\includegraphics[width=6cm]{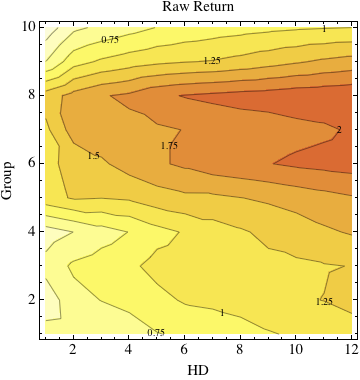}}
		\subfigure[weekly]{\includegraphics[width=6cm]{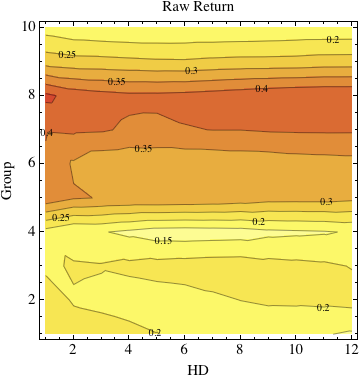}}
		\caption{Raw returns by holding periods and size groups for the KOSPI 200. The x-axis is for holding period and y-axis for market capitalization. Each number means the average return per month and week, respectively. 2.5\% corresponds to red for monthly scale and 0.5\% for weekly scale.}
		\label{FIG_MCAP_Momentum_Return}		
	\end{figure}
	
	The finding in the previous section is cross-checked with the performance of each size group in the KOSPI 200. The raw returns by each size group are given in Fig. \ref{FIG_MCAP_Momentum_Return}. There exists a strong tendency that other size groups are defeated by the groups, S6, S7, and S8, that are equivalent to the top 41-100 companies in market capitalization. Although there exist some variations along with the holding period, the difference is negligible and the tendency is consistent. Even when the total number of size groups is changed to five, the pattern is identical. Other groups are outperformed by the dominance groups, S6, S7, and S8 which have the large overlap with the components of $(100-50)$. Additionally, the tendency is not dependent with time scales of the lookback and holding horizons. In both of weekly and monthly scale, the identical conclusion, that the top 41-100 companies in market capitalization have the stronger performance than other components, is obtained. These returns are all statistically significant.

	When the same test is repeated in other subsets, the identical conclusion is obtained although the results are not presented in the paper. For examples, when the KOSPI 50 is used, the group returns by market capitalization show that large companies corresponding to the S10 for the KOSPI 200 underperform small stocks which are in the S8 for the KOSPI 200. In the case of $(100-50)$, big firms, the S8 for the KOSPI 200, beat small-sized firms, the S6 for the KOSPI 200. The outcomes from $(200-100)$ and $(200-50)$ are similar to the previous results from $(100-50)$ that the large-caps, the S5 for the KOSPI 200, perform better than the small-caps, the S1 for the KOSPI 200. All these results are opposite to the results from $(50)$. Additionally, the outcomes are statistically significant with rejecting the hypothesis that the average return is zero. This significance of the results implies that the market capitalization can play an important role of a predictive criterion for equity price and its predictive direction is changed with respect to the market universe. Moreover, the market capitalization is highly related to the momentum universe shrinkage effect.
		
	However, this conclusion is slightly different with the previous observation \cite{Kim:2012} that reports a simple negative correlation between return and market capitalization. The composition of the KOSPI 200 is one of the reasons for the discrepancy. Although the KOSPI 200 contains small-sized companies, the KOSPI 100, a subset of the KOSPI 200, has severe overlap with the top 100 largest companies in the market and it makes the KOSPI 200 contain higher portions of the big firms than the entire market does. In some senses, many equities used in this paper can be categorized into the large-cap companies among all enlisted stocks. Comparing with the case of all enlisted stocks, our tests on the KOSPI 200 and its related subsets are considered as detail scans on the large company biased universes. When the detail scans on the biased universes are conducted, it is found that the relation between return and size is slightly skewed to the middle-sized equities. If all equities are covered, this skewed correlation is erased and diluted by larger numbers of the equities in the other size groups.
	
	The conclusion of this section is that market capitalization is the one of the main factors which can forecast future performance. However, the correlation between future performance and market capitalization is not as simple as it is known. By changing the market pools, the correlation behaves in different ways. In particular, the top 41-100 stocks in market capitalization perform better than other components in the KOSPI 200. This finding can be explained by the following argument. The size is a barometer for investors' interests and higher cap companies get more attentions which can cause much heavier trades. These attentions make difficult the price move up or down because many investors in the market have watched the largest companies. For examples, analysts tend to write more reports on large companies and this tendency makes the information on the companies disseminate over the market. Relatively smaller but still large-cap companies can move rather easily in any directions. These companies are also relatively good at their businesses because they are qualified to be components of the indexes by the KRX. The outperformance by $(100-50)$ in the shrinkage effect is cross-checked by the size portfolio. The difference of investors' interests can be explained by the liquidity.  In next section, we need to look at the liquidity closely for this reason.

\section{Liquidity and liquidity portfolio}	
	As explained, the liquidity can be a possible answer to the universe shrinkage effect in the momentum returns. To verify it, more tests on the liquidity are needed. Definitely, the liquidity is considered an important market information and it is reported that a negative correlation between past liquidity and future return exists \cite{Lee:2000,Amihud:1986,Amihud:2002,Datar:1998, Hu:1997,Kim:2012}. However, there are many different definitions on the liquidity. Amihud and Mendelson \cite{Amihud:1986} employed bid-ask spread as a proxy of the liquidity. The monthly average of absolute daily return over daily volume in dollar was also suggested for a liquidity measure \cite{Amihud:2002}. Meanwhile, some studies used turnover or fractional turnover \cite{Lee:2000,Datar:1998, Hu:1997}. Since it is difficult to calculate the bid-ask spread from daily time series of price which are the data used in the paper, the fractional volume plays a role of the liquidity proxy in this paper and the liquidity means the fractional volume from now on.
	
	Conclusions from the various literatures on fractional volume, defined by volume divided by outstanding shares also known as turnover rate, are already well-known. In particular, Datar et al. {\cite{Datar:1998} reported that past liquidity has a negative correlation with future return. Lee and Swaminathan \cite{Lee:2000} confirmed Datar's finding and also found that the momentum returns become more outstanding among higher fractional turnover stocks. The similar results are obtained  in the South Korean markets \cite{Kim:2012}. Moreover, in the work by Choi \cite{Choi:2012a} that defined the physical momentum of equities using the fractional volume as mass, the momentum strategies using physical momentum exhibited the stronger performance in absolute return and Sharpe ratio than the traditional momentum strategy although it is governed by the reversion. In these senses, the liquidity not only play a roles of the momentum driving factors but also admits more lucrative strategies.
		
	Based on the results from the previous studies, the trading strategies ranked by the fractional volume are implemented. On the reference day, the equities in the momentum universe are sorted by the past fractional volumes during the estimation period. The most illiquid equities are assigned to the loser group (L1) and the most liquid equities are allocated to the winner group (L10). And then, the L1 is sold and the L10 is bought with the same amount in order to construct the dollar-neutral momentum portfolio. After a given holding period, the portfolio is liquidated by selling the L10 and buying the L1 back.

	\begin{figure}[]
		\subfigure[monthly]{\includegraphics[width=6cm]{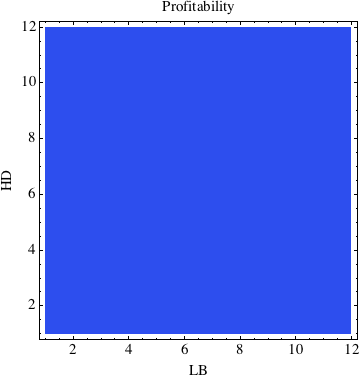}}
		\subfigure[monthly]{\includegraphics[width=6cm]{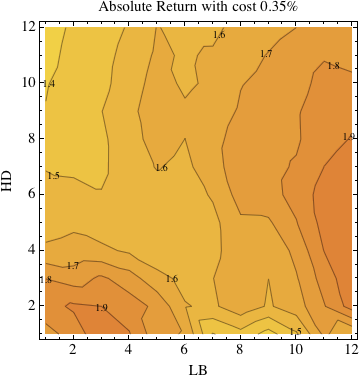}}
		\subfigure[monthly]{\includegraphics[width=6cm]{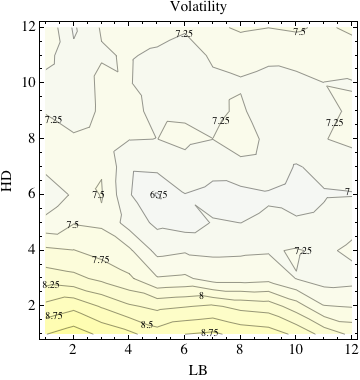}}
		\subfigure[monthly]{\includegraphics[width=6cm]{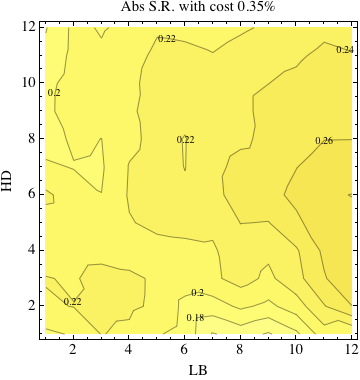}}
		\caption{Statistics for monthly liquidity portfolios in the KOSPI 200. All statistics are in the monthly scale.}
		\label{FIG_LIQ_V_Momentum_Statistics}
	\end{figure}
	
	First of all, the liquidity based strategies are back-tested in the KOSPI 200 and major statistics are given in Fig. \ref{FIG_LIQ_V_Momentum_Statistics}. Profitabilities of the liquidity portfolios exhibit that the small liquidity equities outperform the large liquidity equities on all pairs of estimation and holding periods. This observation is consistent with the well-known results in Datar et al. {\cite{Datar:1998}, Lee and Swaminathan \cite{Lee:2000}, and other works \cite{Hu:1997,Kim:2012}. Returns by the liquidity portfolios are also more consistent along with lookback and holding period parameters than the cumulative return-based momentum strategies in the KOSPI 200. Over all parameter sets, the returns by the liquidity portfolios vary approximately between 1.4\% and 1.9\% after subtracting the transaction cost of 35 bps per each of winner and loser baskets. Moreover, the returns of the liquidity portfolios are statistically significant. Meanwhile, the cumulative return based momentum strategies are in the wider range between -0.5\% and 2\%. Since the dependence on the lookback and holding periods for the liquidity portfolio is smaller, the expected returns are under smaller risks of possible abrupt change on the optimal strategy. The volatility graphs also support this stable performance of the liquidity portfolios. Though the cumulative return-based momentum strategy has volatilities of roughly minimum 8\% and maximum 12\%, volatilities of the liquidity portfolios have only 1.5\% difference between minimum of approximately 7.25\% and maximum of 8.75\%. The best Sharpe ratio from the traditional momentum strategies is around 0.2 but the liquidity portfolios have greater Sharpe ratios than 0.2 at almost all lookback and holding sites.

	\begin{figure}[]
		\begin{center}
		\subfigure[$(100)$]{\includegraphics[width=3.6cm]{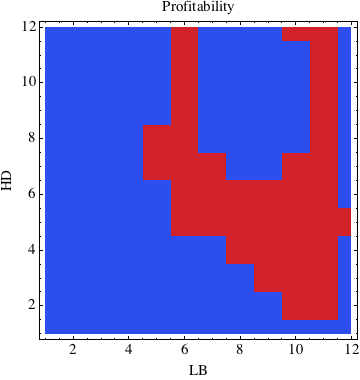}}
		\subfigure[$(50)$]{\includegraphics[width=3.6cm]{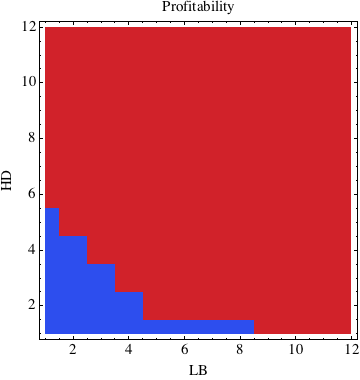}}
		\subfigure[$(200-50)$]{\includegraphics[width=3.6cm]{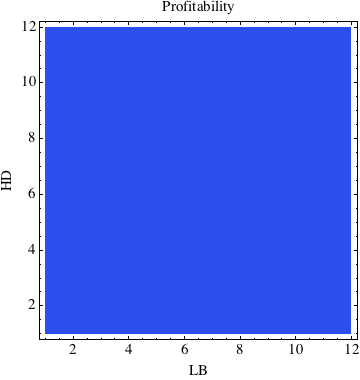}}
		\subfigure[$(200-100)$]{\includegraphics[width=3.6cm]{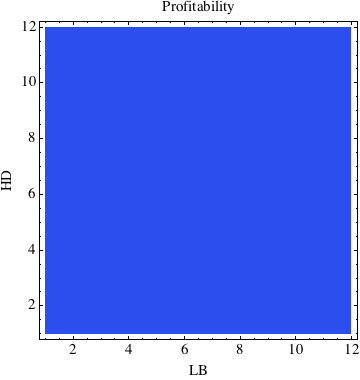}}
		\subfigure[$(200-100+50)$]{\includegraphics[width=3.6cm]{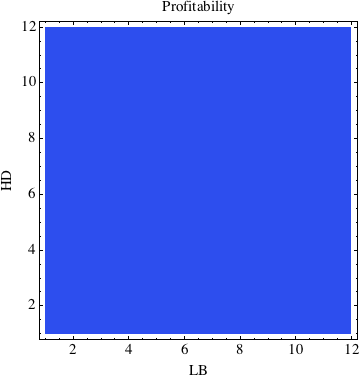}}
		\subfigure[$(100-50)$]{\includegraphics[width=3.6cm]{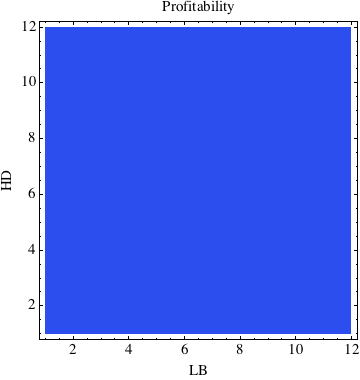}}
		\end{center}
		\caption{Profitabilities of monthly liquidity portfolios in the subuniverses.}
		\label{FIG_LIQ_V_Momentum_Profitability}
	\end{figure}
	
	When the market pool is changed to one of the subuniverses of the KOSPI 200, profitabilities of the liquidity portfolios in the subsets, given in Fig. \ref{FIG_LIQ_V_Momentum_Profitability}, are different with those of the KOSPI 200. In particular, the universes except for the KOSPI 100 and the KOSPI 50 have negative raw returns indicating that past low volume stocks outperform past high volume stocks, similar to the KOSPI 200. In addition to that, as the market universe includes the KOSPI 50 which has the higher market capitalization stocks, the area for positive raw returns in the profitability graphs is widened. It means that a correlation between past liquidity and future return is changed to be positive while many large firms exist in the universe. In the KOSPI 100 and the KOSPI 50, the liquid equities tend to outperform the illiquid equities and this outcome is opposite to the observation from the KOSPI 200 and other subsets in which there exists the negative correlation between past liquidity and future return.
	
	The return results with the transaction costs of 35 bps for each subuniverse are shown in Fig. \ref{FIG_LIQ_V_Sub_Momentum_Return}. Same to the return-based momentum, the liquidity portfolios also have the statistically significant dependence on existence of the KOSPI 50 components, i.e. $(100)$ and $(50)$ have weaker returns than $(200)$. When it is excluded from the universe, the performance is improved and the examples are $(200-50)$, $(200-100)$, and $(100-50)$. A comparison between $(200-100)$ and $(200-100+50)$ also shows that the existence of the KOSPI 50 components hinders the portfolio performance. It is supported that the momentum universe shrinkage effect is still persistent even with different ranking criteria.
	
	\begin{figure}[]
		\subfigure[$(100)$]{\includegraphics[width=6cm]{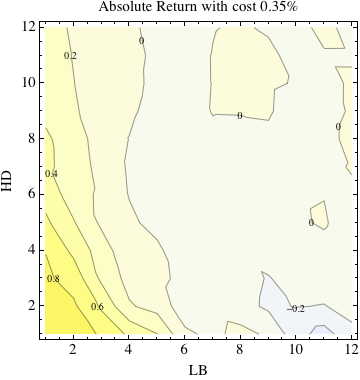}}
		\subfigure[$(50)$]{\includegraphics[width=6cm]{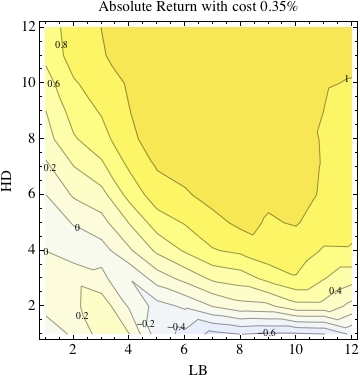}}
		\subfigure[$(200-50)$]{\includegraphics[width=6cm]{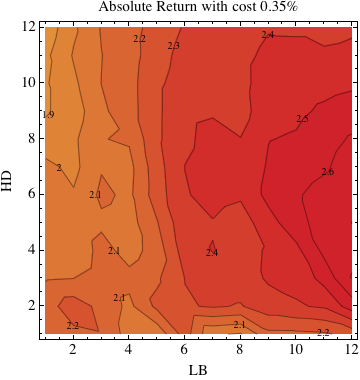}}
		\subfigure[$(200-100)$]{\includegraphics[width=6cm]{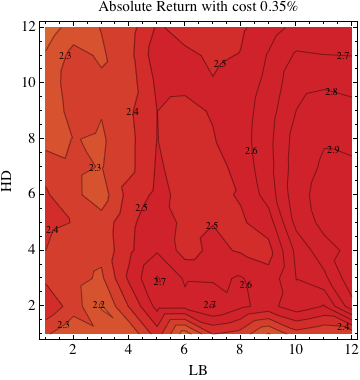}}
		\subfigure[$(200-100+50)$]{\includegraphics[width=6cm]{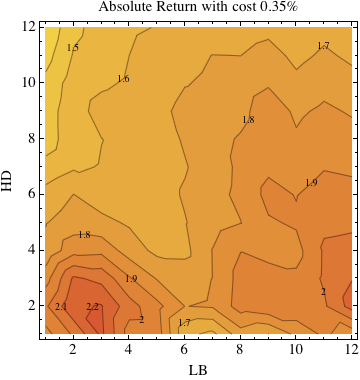}}
		\subfigure[$(100-50)$]{\includegraphics[width=6cm]{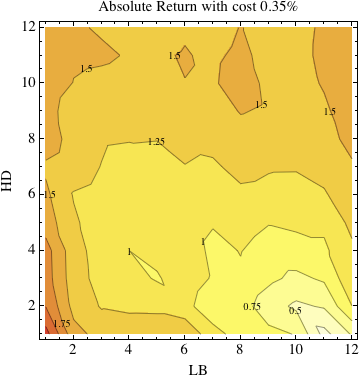}}
		\caption{Returns of monthly liquidity porfolios in the subuniverses. All returns are in the monthly scale.}
		\label{FIG_LIQ_V_Sub_Momentum_Return}		
	\end{figure}
		
	The Sharpe ratios in Fig. \ref{FIG_LIQ_V_Sub_Momentum_SR} also indicate that the market universes containing the KOSPI 50 components underperform the non-KOSPI 50 universes. The Sharpe ratios from $(200-50)$ and $(200-100)$ are better than those of $(200)$. Moreover, $(100-50)$ has larger Sharpe ratios than $(100)$ and $(50)$. The only difference is that $(200-100+50)$ has the better Sharpe ratios than $(200)$ but the smaller Sharpe ratio than $(200-100)$. In the case of the cumulative return-based momentum, the Sharpe ratios of $(200-100+50)$ are smaller than those of $(200)$ and $(200-100)$. Ignoring a few exceptions, the general tendency of returns and Sharpe ratios is identical to that of the cumulative return-based momentum and the liquidity criterion can improve performance and stability of the portfolio.
	
	\begin{figure}[]
		\subfigure[$(100)$]{\includegraphics[width=6cm]{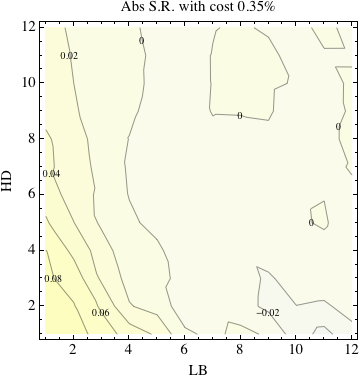}}
		\subfigure[$(50)$]{\includegraphics[width=6cm]{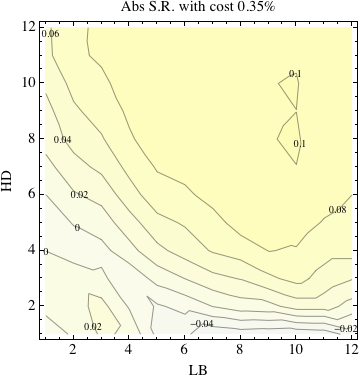}}
		\subfigure[$(200-50)$]{\includegraphics[width=6cm]{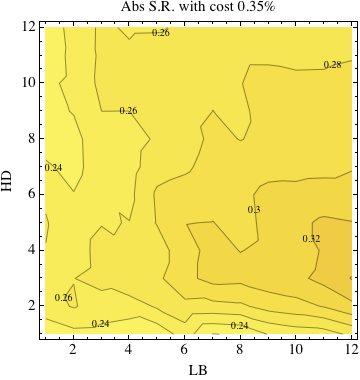}}
		\subfigure[$(200-100)$]{\includegraphics[width=6cm]{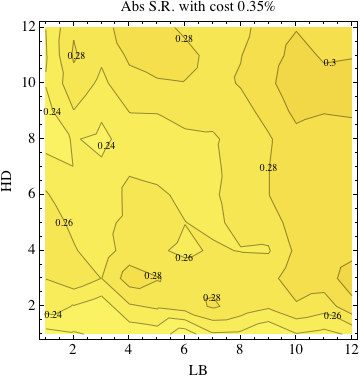}}
		\subfigure[$(200-100+50)$]{\includegraphics[width=6cm]{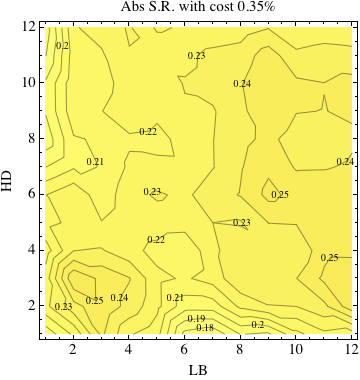}}
		\subfigure[$(100-50)$]{\includegraphics[width=6cm]{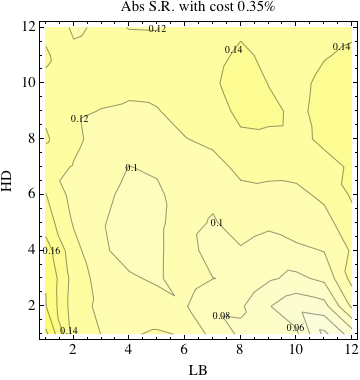}}
		\caption{Sharpe ratios of monthly liquidity portfolios in the subuniverses. All relative Sharpe ratios are in the monthly scale.}
		\label{FIG_LIQ_V_Sub_Momentum_SR}		
	\end{figure}
	
	From the observations, it is evident that the liquidity portfolios also have the dependence on the momentum universe shrinkage. In particular, the KOSPI 50 components are not useful for increasing performance of the liquidity portfolios similar to the previous criteria such as cumulative return and market capitalization. The universe effect is still persistent regardless of the ranking criteria. The returns and Sharpe ratios of the liquidity portfolios in $(200-50)$ and $(200-100)$ indicate that the negative correlation between liquidity and return becomes stronger among the small-caps. Since the liquidity is provided by the market participants, we need to focus on the behavior of investors in the market.
	
\section{Investor groups and transaction portfolio}
	The financial markets have various categories of market participants. Investors have different strategies, characteristics, amounts of capitals, expectations on future movement of price, time horizons for investment, and risk tolerances. The market participants are divided into two broad classification groups, individual and institutional investors. The individuals are also called personal investors who don't work for the specific financial firms. They generally invest some portions of their own household incomes and assets. The individual investors' behavior and their performance are covered by various literatures \cite{Schlarbaum:1978a,Schlarbaum:1978b,Odean:1998a,Barber:2000}. Meanwhile, the institutional investors are usually mutual funds, pension funds, hedge funds, and banks. They tend to have larger amounts of assets under management than the individuals do. Similar to the individual investors' case, the institutional investors are also concerns of academics and diverse studies on their trading patterns and earnings are conducted \cite{Grinblatt:1995,Carhart:1997}. With consideration on their funding sources, these institutional investors are classified into domestic and foreign institutions. In particular, this classification becomes much clearer in emerging markets. Based on origins and affiliations, investors in the emerging markets are largely categorized into three different groups such as individual, institutional, and foreign investors.
	 
	In particular, the foreign investors in emerging markets have given serious market impacts and are sometime blamed because their trading patterns tend to increase volatility of markets. It is believed that they have more accurate expectation models, hold large portion of shares in the local exchanges, and get more profits from emerging markets. They also exhibit the positively correlated trading pattern with global and domestic equity returns \cite{Richards:2005}. The foreign investors generate serious market impacts because they can trade assets massively with their huge amounts of capitals. Since their impacts to the emerging markets are important, there are many financial literatures on foreign investors in various emerging markets including Asian markets \cite{Richards:2005,Li:2011,Chen:2005} and the South Korean markets \cite{Richards:2005, Kim:2002,Ko:2007}.
	
	In the South Korean markets, foreign investors are registered and their transactions are recorded in terms of daily volume and amount of transaction in cash. From the transactional information, ownership by foreign investors per equity is also calculated by the KRX. The transaction information of institutional investors for all enlisted equities is also provided by the KRX. The daily transaction data by individuals are inferred by the opposite to the sum of institutional and foreign trades because it is assumed that the relation of $T_{ind}+T_{ins}+T_{for}=0$ holds where $T_{ind}, T_{ins},$ and $T_{for}$ are trades by individual, institutional, and foreign investors, respectively. With the transactional information, a trading pattern by each group can be analyzed.
	
	For the analysis, a trading strategy which uses the transaction information as a ranking criterion is implemented. Similar to the liquidity portfolio, the criterion is defined by fractional trading volume, which is a ratio of daily volume to total outstanding shares. For examples, the winner portfolio (T10) is constructed from the equities which have the largest net fractional buying by a given investor group and the loser group (T1) is decided by the most selling-offs of the same group during lookback periods. The construction of the transaction portfolio encodes trading patterns and viewpoints of that group. The profitability exhibits whether or not they make correct expectations and the magnitude of the portfolio return decides how strong a correlation between their past trades and future returns is. If the portfolio shows a positive expected return, equities which a given group has bought in the past tend to perform better in the future. In the opposite case, the transactional data has a negative correlation with the future price. In any cases, the transactional data by investor groups can have the predictive power on future price movements and test how robust the influence by the group is if there exists the correlation regardless of its sign. After finding the correlation, dependence on the universe shrinkage will be tested.
	
	\begin{figure}[]
		\subfigure[individual investors]{
			\includegraphics[width=3.6cm]{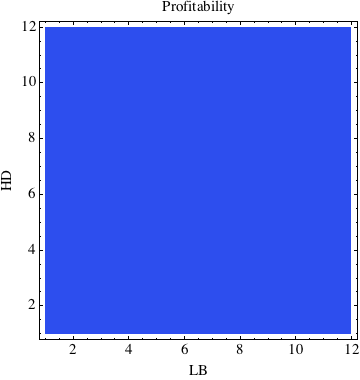}
			\includegraphics[width=3.6cm]{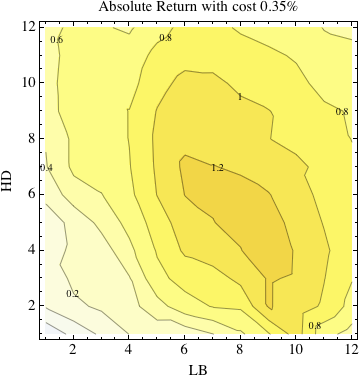}
			\includegraphics[width=3.6cm]{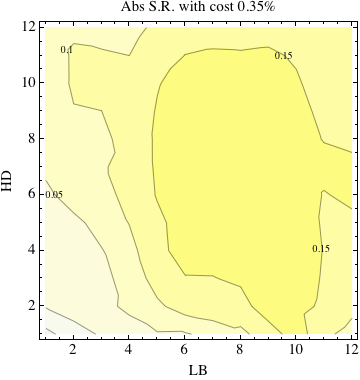}
		}
		\subfigure[institutional investors]{
			\includegraphics[width=3.6cm]{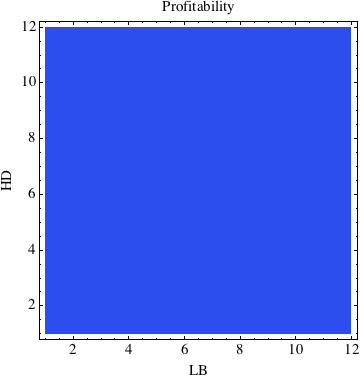}
			\includegraphics[width=3.6cm]{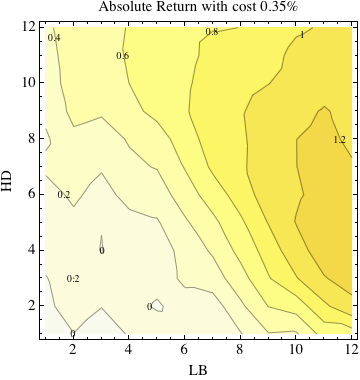}
			\includegraphics[width=3.6cm]{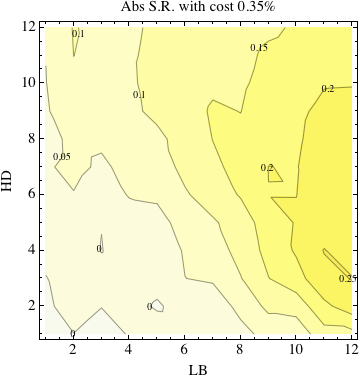}
		}
		\subfigure[foreign investors]{
			\includegraphics[width=3.6cm]{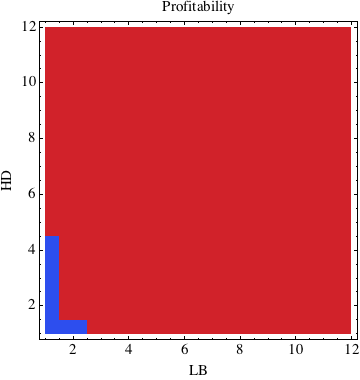}
			\includegraphics[width=3.6cm]{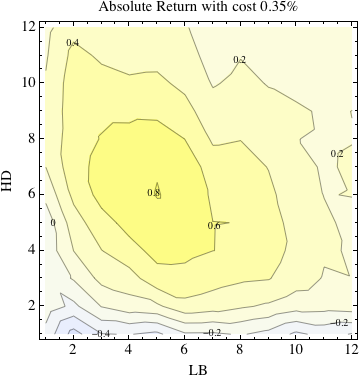}
			\includegraphics[width=3.6cm]{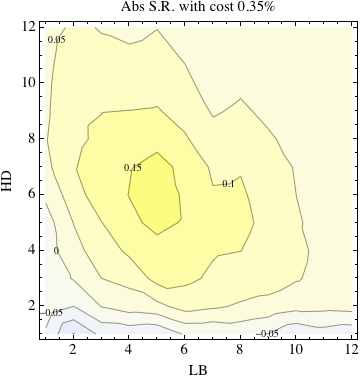}
		}
		\caption{Profitabilities, returns, and Sharpe ratios of monthly investors' transaction portfolios in the KOPSI 200. All returns and Sharpe ratios are in the monthly scale. 2.5\% corresponds to the red and -2.5\% for blue in return graphs.}
		\label{FIG_TRST_Momentum_Return_SR_KP200}		
	\end{figure}
	
	The profitabilities and absolute returns for investor group's transaction based strategies in the KOSPI 200 are found in Fig. \ref{FIG_TRST_Momentum_Return_SR_KP200}. Many implemented returns are statistically significant. Each investor group has a different trading pattern and gives diverse market impacts. For examples, considering the profitabilities of the transaction portfolios by individuals, equities which individual investors have bought tend to underperform equities which they have sold. The maximum is located near at the 8-month lookback and 6-month holding strategy. The stocks bought by institutional investors over recent 12 months are likely to perform poorer during the same holding period with that of the individuals. The trading patterns by individual and institutional investors are negatively correlated with future returns in the KOSPI 200. Meanwhile, foreign investors are different with other groups. Stocks purchased by the foreigners outperform stocks they sold in the past. In some sense, they are good predictors because the purchased equities perform well in near future. The foreign investors have the relatively shorter estimation period. Moreover, its maximum is much smaller than those of the portfolios by individual and institutional investors. It is supposed that the foreign investors change their expectation and strategies more frequently than other investors because of the shorter lookback period and smaller expected return for the best strategy. The Sharpe ratios by each investors are similar to the return results.
	
	\begin{figure}[]
	\begin{center}
		\subfigure[$(100)$]{\includegraphics[width=3.6cm]{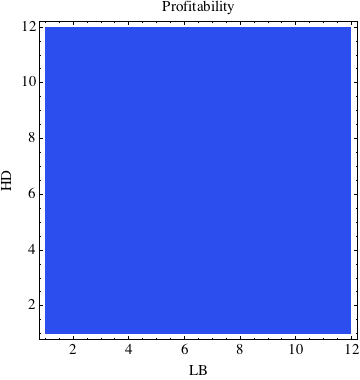}}
		\subfigure[$(50)$]{\includegraphics[width=3.6cm]{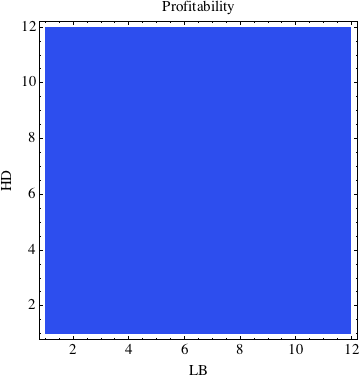}}
		\subfigure[$(200-50)$]{\includegraphics[width=3.6cm]{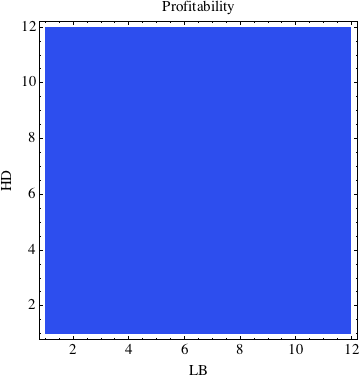}}
		\subfigure[$(200-100)$]{\includegraphics[width=3.6cm]{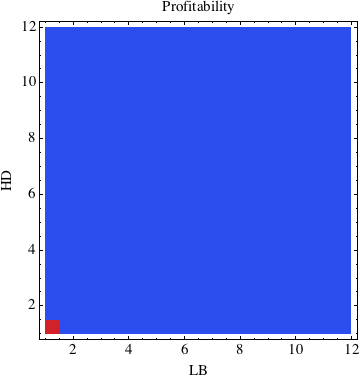}}
		\subfigure[$(200-100+50)$]{\includegraphics[width=3.6cm]{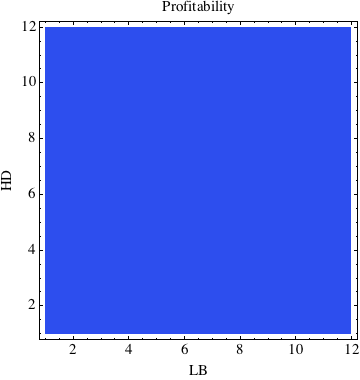}}
		\subfigure[$(100-50)$]{\includegraphics[width=3.6cm]{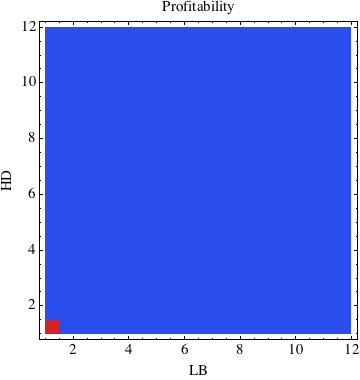}}
	\end{center}
		\caption{Profitabilities of monthly transaction portfolios by individual investors in the subuniverses.}
		\label{FIG_TRST_Momentum_Profitability_IND}		
	\end{figure}
	
	\begin{figure}[]
		\subfigure[$(100)$]{\includegraphics[width=6cm]{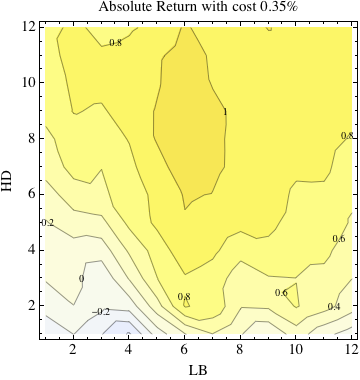}}
		\subfigure[$(50)$]{\includegraphics[width=6cm]{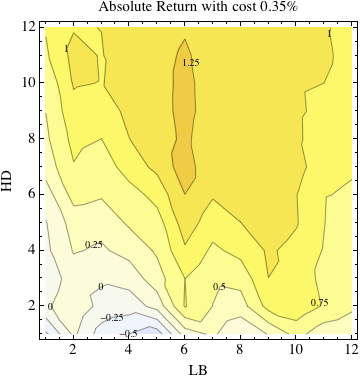}}
		\subfigure[$(200-50)$]{\includegraphics[width=6cm]{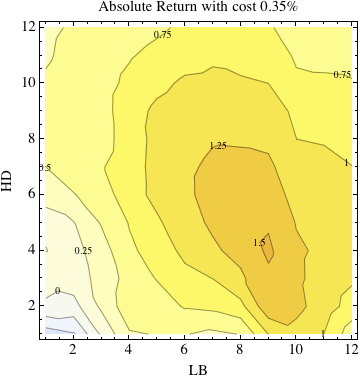}}
		\subfigure[$(200-100)$]{\includegraphics[width=6cm]{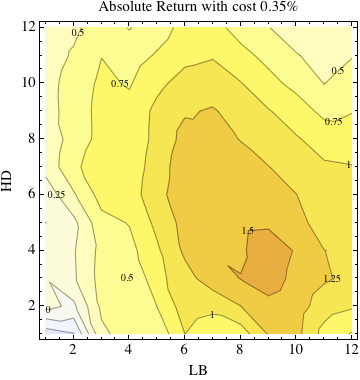}}
		\subfigure[$(200-100+50)$]{\includegraphics[width=6cm]{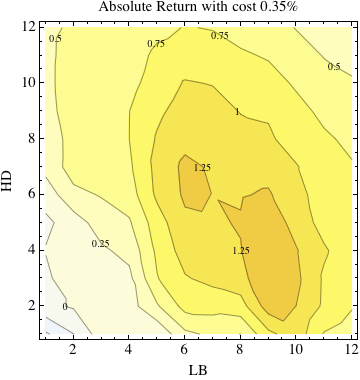}}
		\subfigure[$(100-50)$]{\includegraphics[width=6cm]{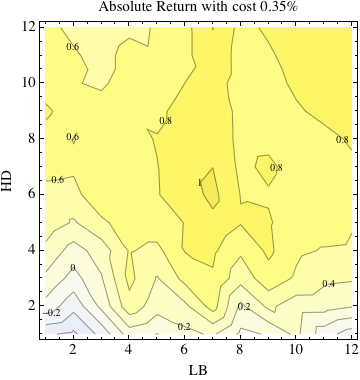}}
		\caption{Returns of monthly transaction portfolios by individual investors in the subuniverses. All returns are in the monthly scale.}
		\label{FIG_TRST_Momentum_Return_IND}		
	\end{figure}
	
	\begin{figure}[]
		\subfigure[$(100)$]{\includegraphics[width=6cm]{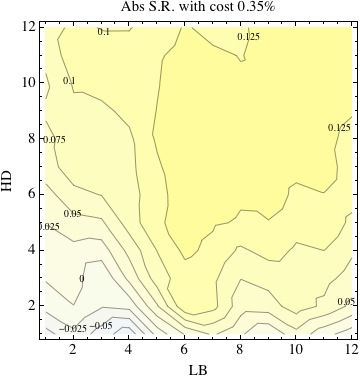}}
		\subfigure[$(50)$]{\includegraphics[width=6cm]{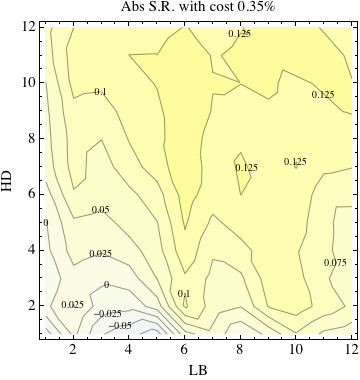}}
		\subfigure[$(200-50)$]{\includegraphics[width=6cm]{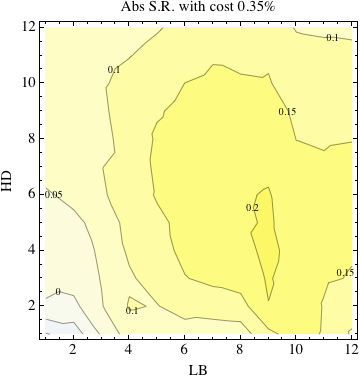}}
		\subfigure[$(200-100)$]{\includegraphics[width=6cm]{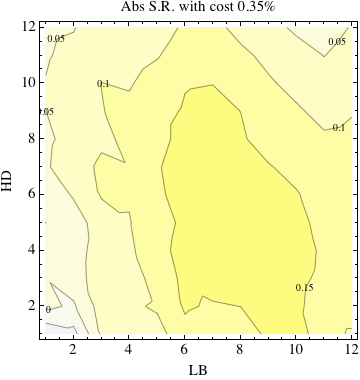}}
		\subfigure[$(200-100+50)$]{\includegraphics[width=6cm]{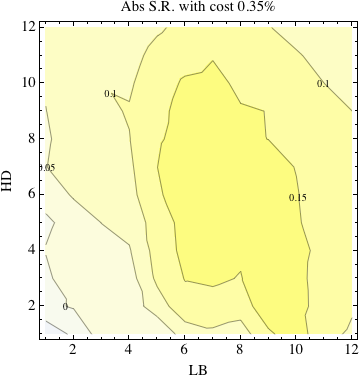}}
		\subfigure[$(100-50)$]{\includegraphics[width=6cm]{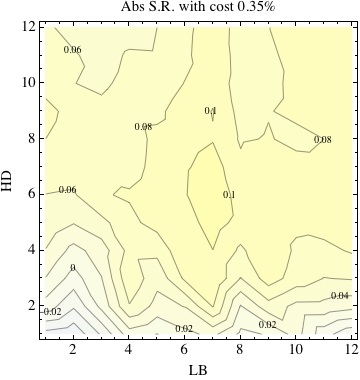}}
		\caption{Sharpe ratios of monthly transaction portfolios by individual investors in the subuniverses. All Sharpe ratios are in the monthly scale.}
		\label{FIG_TRST_Momentum_SR_IND}		
	\end{figure}
		
	The behaviors of individual investors are different in the subuniverses. According to Fig. \ref{FIG_TRST_Momentum_Profitability_IND}, it is apparent that they are all contrarian, identical to the KOSPI 200 case. A negative correlation between individual investors' trading history and future return exists in all subuniverses. These facts are consistent with the prior studies by Odean \cite{Odean:1998a} and Barber and Odean \cite{Barber:2000} reporting that the individual investors get losses from their trades and are reluctant to sell when their accounts are damaged. It is also well-matched with the conclusion of Ko et al. \cite{Ko:2007}. They reported that the equities which both of institutional and foreign investors purchase, i.e. the individuals sell, outperform significantly the equities sold by both of institutional and foreign investors. Similar to the conclusion of these two papers, the individuals sell equities before prices are going up and hold stock positions when prices are down. In Fig. \ref{FIG_TRST_Momentum_Return_IND}, it is obvious that their transaction becomes more meaningful in $(200-50)$, $(200-100)$, $(200-100+50)$, and $(50)$. In particular, the performance in $(200-50)$ and $(200-100)$ shows that the magnitude of returns for small- and middle-size companies become larger than other subuniverses. However, the results from $(200-100+50)$ and $(50)$ also tells that returns of large companies are inversely influenced by individual investors' trades. These seem contradictory but the results from $(50)$, $(100)$ and $(100-50)$ can tell more underlying stories. Opposite to the results in the cumulative return-based momentum strategies, $(100-50)$ underperforms $(100)$ in cases of the transaction portfolios by the individuals and this fact can solve the contradiction. Additionally, $(50)$ outperforms $(100)$. The individual investors give more impacts on large and small companies not on the mid-cap equities. In other words, their trades become extreme in market capitalization. It is also found that the Sharpe ratios in Fig. \ref{FIG_TRST_Momentum_SR_IND} show the similar patterns with the return case although some minor differences are found in the comparisons between $(100)$ and $(50)$ and between $(200-50)$ and $(200-100)$.
	
	\begin{figure}[]
	\begin{center}
		\subfigure[$(100)$]{\includegraphics[width=3.6cm]{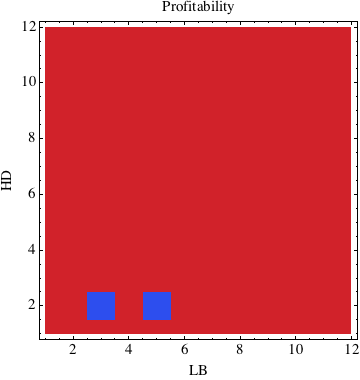}}
		\subfigure[$(50)$]{\includegraphics[width=3.6cm]{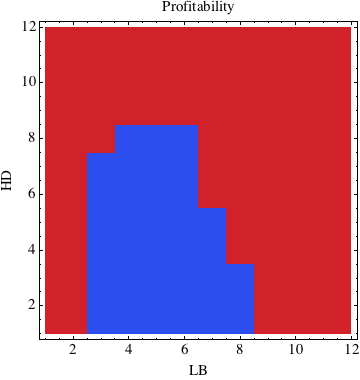}}
		\subfigure[$(200-50)$]{\includegraphics[width=3.6cm]{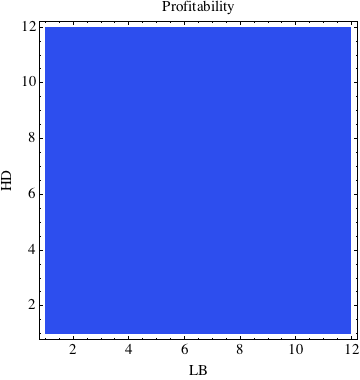}}
		\subfigure[$(200-100)$]{\includegraphics[width=3.6cm]{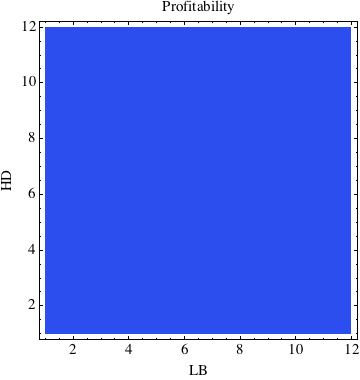}}
		\subfigure[$(200-100+50)$]{\includegraphics[width=3.6cm]{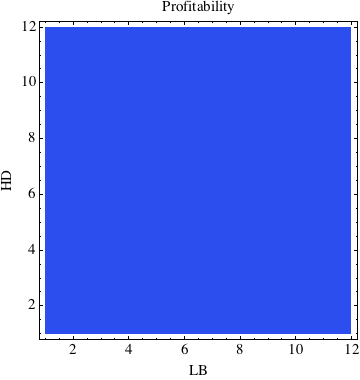}}
		\subfigure[$(100-50)$]{\includegraphics[width=3.6cm]{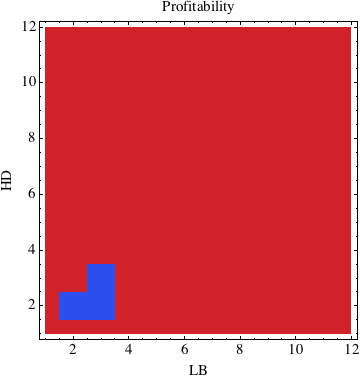}}
	\end{center}
		\caption{Profitabilites of monthly transaction portfolios by institutional investors in the subuniverses.}
		\label{FIG_TRST_Momentum_Profitability_INS}		
	\end{figure}
	
	\begin{figure}[]
		\subfigure[$(100)$]{\includegraphics[width=6cm]{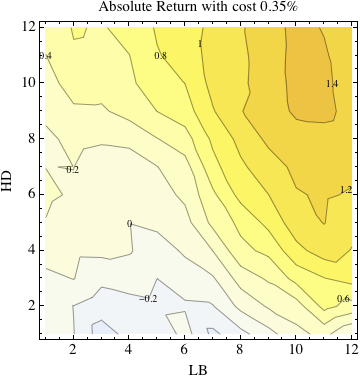}}
		\subfigure[$(50)$]{\includegraphics[width=6cm]{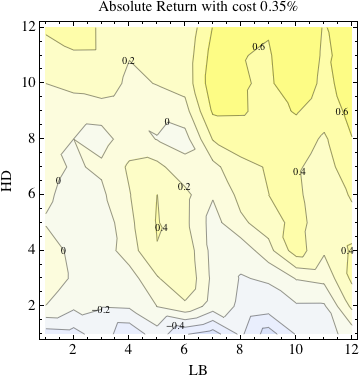}}
		\subfigure[$(200-50)$]{\includegraphics[width=6cm]{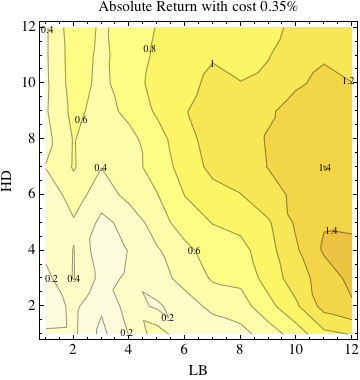}}
		\subfigure[$(200-100)$]{\includegraphics[width=6cm]{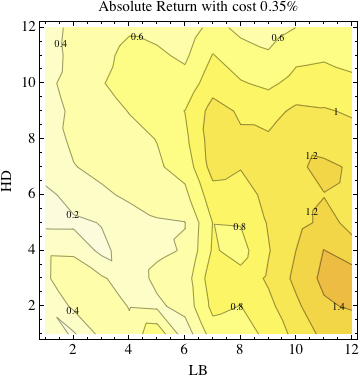}}
		\subfigure[$(200-100+50)$]{\includegraphics[width=6cm]{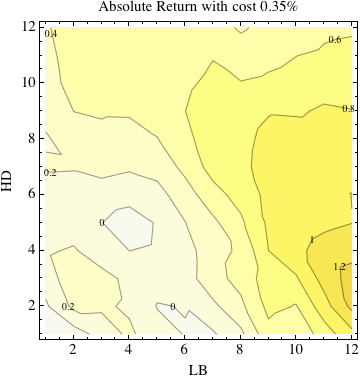}}
		\subfigure[$(100-50)$]{\includegraphics[width=6cm]{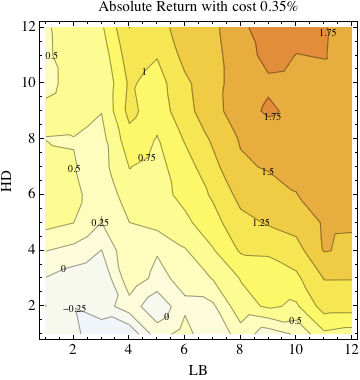}}
		\caption{Returns of monthly transaction portfolios by institutional investors in the subuniverses. All returns are in the monthly scale.}
		\label{FIG_TRST_Momentum_Return_INS}		
	\end{figure}
	
	\begin{figure}[]
		\subfigure[$(100)$]{\includegraphics[width=6cm]{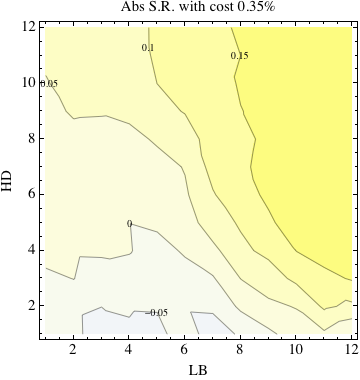}}
		\subfigure[$(50)$]{\includegraphics[width=6cm]{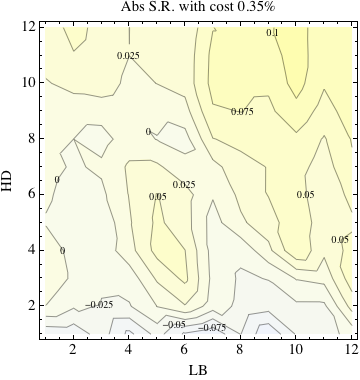}}
		\subfigure[$(200-50)$]{\includegraphics[width=6cm]{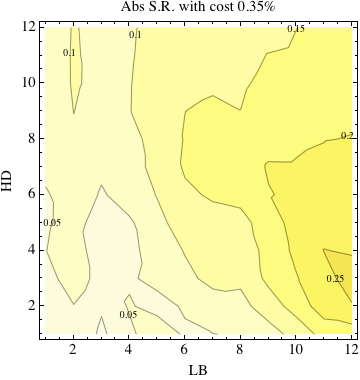}}
		\subfigure[$(200-100)$]{\includegraphics[width=6cm]{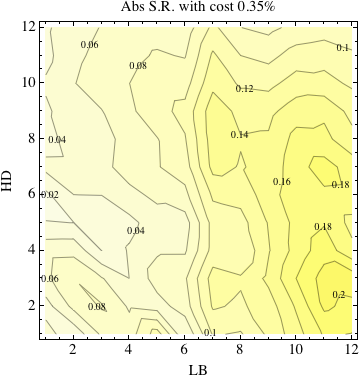}}
		\subfigure[$(200-100+50)$]{\includegraphics[width=6cm]{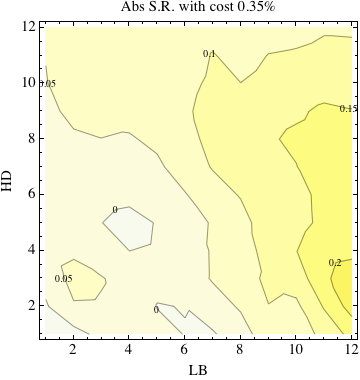}}
		\subfigure[$(100-50)$]{\includegraphics[width=6cm]{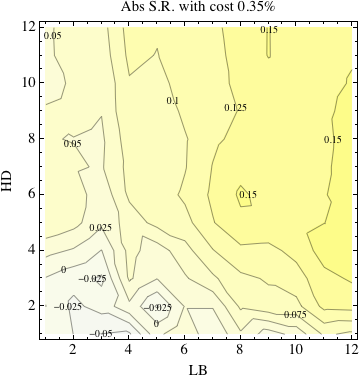}}
		\caption{Sharpe ratios of monthly transaction portfolios by institutional investors in the subuniverses. All Sharpe ratios are in the monthly scale.}
		\label{FIG_TRST_Momentum_SR_INS}		
	\end{figure}
	
	In Fig. \ref{FIG_TRST_Momentum_Profitability_INS}, it is found that the institutional investors have the similar trading patterns with the individual investors except for $(100)$, $(50)$ and $(100-50)$. In those exceptional universes, the institutional investors are good predictors because the equities bought by them outperform the equities sold by them. They are bad forecasters for the small caps and this is supported by the results from $(200-50)$ and $(200-100)$. When the profitability graphs are combined with Fig. \ref{FIG_TRST_Momentum_Return_INS}, their negative correlation becomes more significant in universes containing smaller companies such as $(200-50)$ and $(200-100)$ and a positive correlation becomes stronger in $(100)$ and $(100-50)$. The strong performance of $(100)$ and $(100-50)$ also shows that the institutional investors are good at forecasting in mid-cap companies. They are not astute with the KOSPI 50 components. The weaker performance of the portfolio in $(50)$ is another example and a comparison between $(200-100)$ and $(200-100+50)$ also guarantees that the KOSPI 50 components are not well-predictable by the institutional investors. The strong returns in $(100-50)$ and poor returns in $(50)$ are related to the momentum universe shrinkage that the non-KOSPI 50 related universes have the stronger momentum performance. It is also compatible with the observation in Kim et al. \cite{Ko:2007} that the average market capitalization of what institutional investors trade is relatively smaller than that of foreign investors. The Sharpe ratios are given in Fig. \ref{FIG_TRST_Momentum_SR_INS} which reports the consistent conclusion although there exists a minor difference in $(100-50)$.
	
	\begin{figure}[]
	\begin{center}
		\subfigure[$(100)$]{\includegraphics[width=3.6cm]{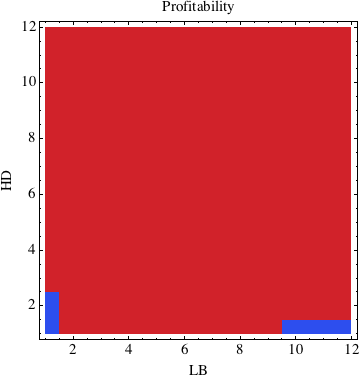}}
		\subfigure[$(50)$]{\includegraphics[width=3.6cm]{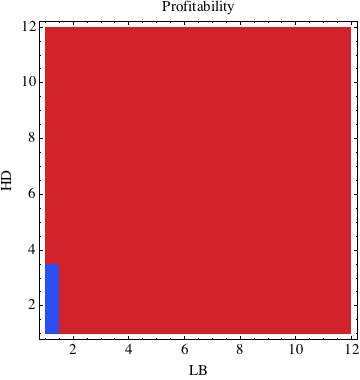}}
		\subfigure[$(200-50)$]{\includegraphics[width=3.6cm]{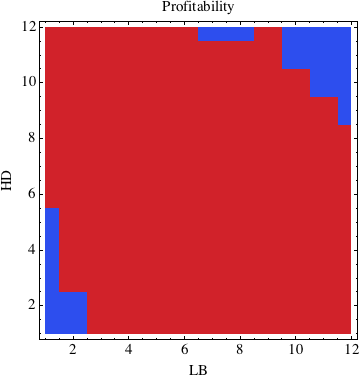}}
		\subfigure[$(200-100)$]{\includegraphics[width=3.6cm]{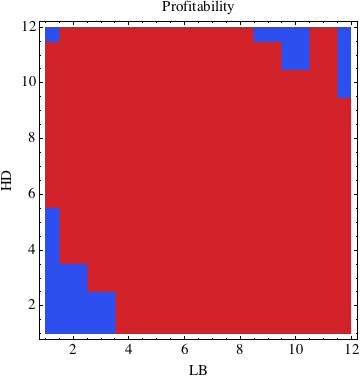}}
		\subfigure[$(200-100+50)$]{\includegraphics[width=3.6cm]{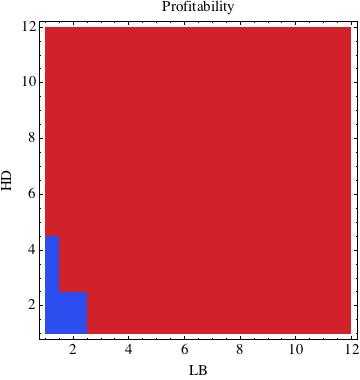}}
		\subfigure[$(100-50)$]{\includegraphics[width=3.6cm]{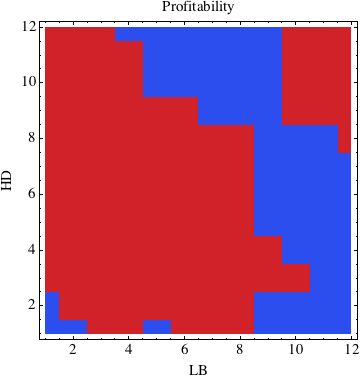}}
	\end{center}
		\caption{Profitabilities of monthly transaction portfolios by foreign investors in the subuniverses.}
		\label{FIG_TRST_Momentum_Profitability_FOR}		
	\end{figure}
	
	\begin{figure}[]
		\subfigure[$(100)$]{\includegraphics[width=6cm]{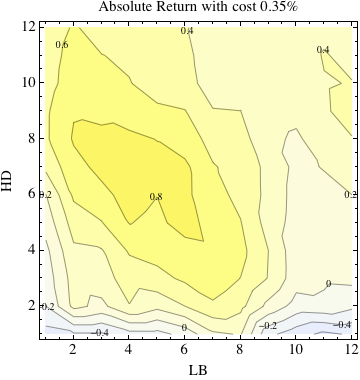}}
		\subfigure[$(50)$]{\includegraphics[width=6cm]{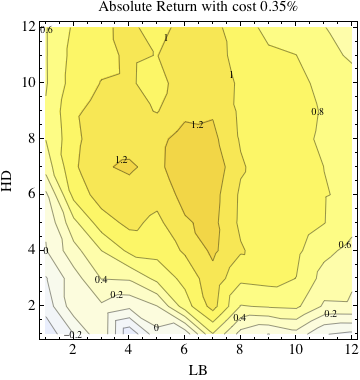}}
		\subfigure[$(200-50)$]{\includegraphics[width=6cm]{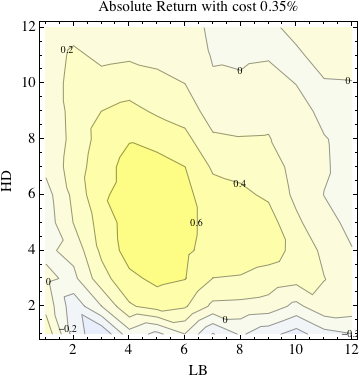}}
		\subfigure[$(200-100)$]{\includegraphics[width=6cm]{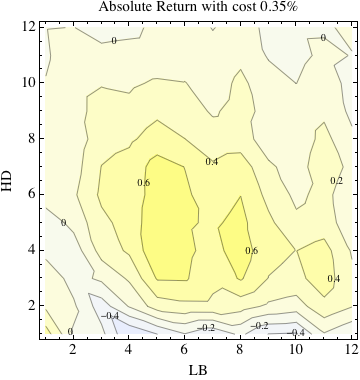}}
		\subfigure[$(200-100+50)$]{\includegraphics[width=6cm]{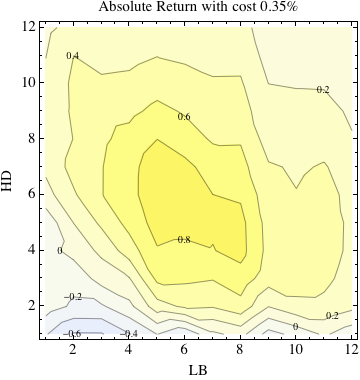}}
		\subfigure[$(100-50)$]{\includegraphics[width=6cm]{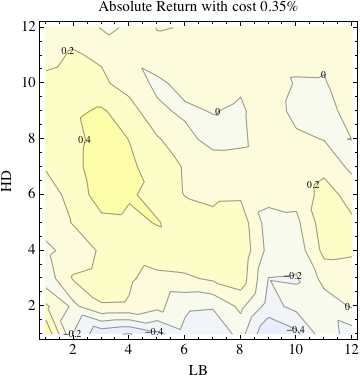}}
		\caption{Returns of monthly transaction portfolios by foreign investors in the subuniverses. All returns are in the monthly scale.}
		\label{FIG_TRST_Momentum_Return_FOR}		
	\end{figure}
	
	\begin{figure}[]
		\subfigure[$(100)$]{\includegraphics[width=6cm]{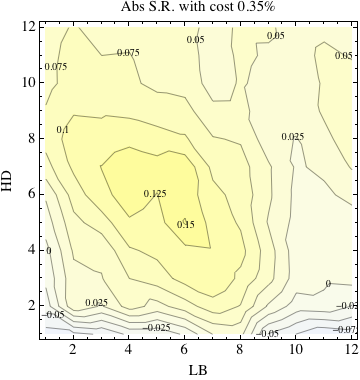}}
		\subfigure[$(50)$]{\includegraphics[width=6cm]{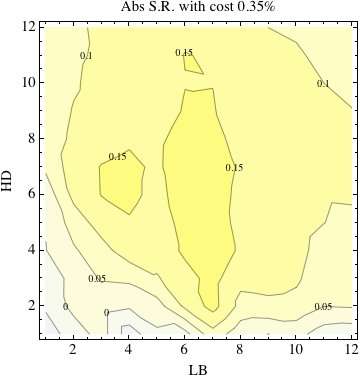}}
		\subfigure[$(200-50)$]{\includegraphics[width=6cm]{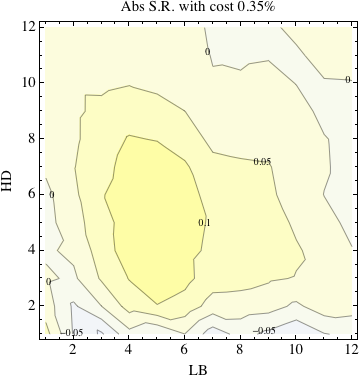}}
		\subfigure[$(200-100)$]{\includegraphics[width=6cm]{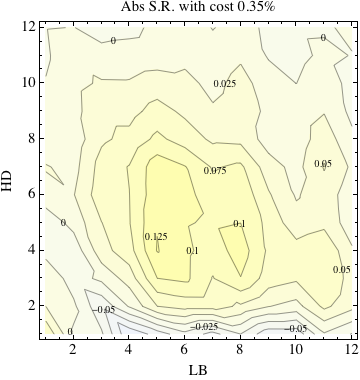}}
		\subfigure[$(200-100+50)$]{\includegraphics[width=6cm]{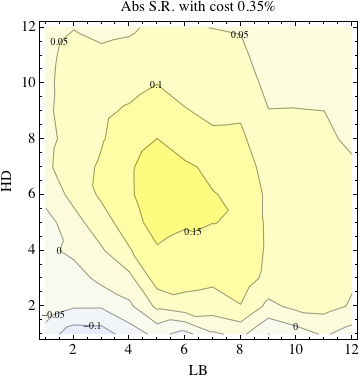}}
		\subfigure[$(100-50)$]{\includegraphics[width=6cm]{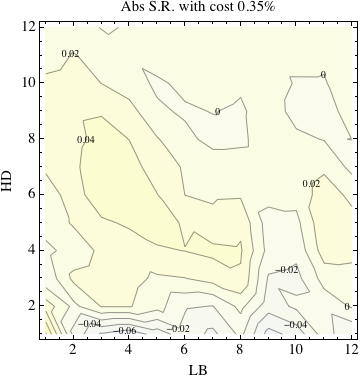}}
		\caption{Sharpe ratios of monthly transaction portfolios by foreign investors in the subuniverses. All Sharpe ratios are in the monthly scale.}
		\label{FIG_TRST_Momentum_SR_FOR}		
	\end{figure}
	
	Unlike the individual and institutional investors, the foreign investors are good forecasters in any market universes because the equities bought by them outperform what they sold. In particular, the profitability graphs in Fig. \ref{FIG_TRST_Momentum_Profitability_FOR} show that their predictability grows in $(100)$, $(50)$, and $(200-100+50)$ which have the wider areas of outperformance. Opposite to the other investors, the positive profitability zones become larger as $(50)$ is included into the universes. It is also well-matched to the observations in Ko et al. \cite{Ko:2007} that foreign investors in the South Korean markets prefer large-sized companies. The return graphs, given in Fig. \ref{FIG_TRST_Momentum_Return_FOR}, also support this conclusion. The returns of the foreign transaction portfolios in $(50)$ are largest and a comparison between $(100-50)$ and $(100)$ shows that inclusion of the KOSPI 50 components increases returns of the portfolios in the case of foreign investors. Additionally, $(200-100+50)$ also outperforms $(200-100)$. As $(50)$ is included, the foreign investors tend to be good predictors in their trading, i.e. the equities with their purchases perform better than what they bought in the past. The Sharpe ratios shown in Fig. \ref{FIG_TRST_Momentum_SR_FOR} also provide the same conclusion with the return graphs. The fact, that they are only good at forecasting future prices of the KOSPI 50 components, can give a part of the answer to the market shrinkage effect. The reason why they prefer large firms is following. Since the foreign investors are limited in acquiring the local information, they tend to focus on relatively well-known companies which generally have the larger market capitalization. Additionally, it is believed that they have the better pricing models and this is why the large firms they have bought outperform. 
	
	According to the returns by the transaction portfolios, foreign investors mostly have a sharp edge on prediction of prices for the KOSPI 50 components and institutional investors are good forecasters mainly for $(100-50)$. The individual investors are bad forecasters in all universes, particularly in smaller market capitalization universes but the magnitude of their market impacts in the subuniverses for small-sized companies is not negligible. Each group has his/her main playgrounds and this fact seems to be related to the market shrinkage effect. The reason of this preference is following. Although the foreign investors have better strategies and larger assets, their information on the firms enlisted in local markets of emerging countries is not as accurate as that of domestic investors, mostly domestic institutional investors. This is why they focus much more on the largest companies which are the members of the KOSPI 50. The institutional investors have enough information but tend to give less market impacts on the largest equities than the foreign investors do. Since they have less assets and more crude strategies than the foreign investors, they decide not to compete with the foreigners and choose $(100-50)$ as their playground which has still large-cap companies but is less informative to the foreigners. Meanwhile, the individual investors find their edges in the small-sized companies not to compete with other two groups although their profits are not promised.

\section{Conclusion}
	In this paper, it is found that the momentum returns are not homogeneous in the subsets of the momentum universe in the South Korean markets and the market shrinkage effect had been not reported in any previous studies. In some universes including the KOSPI 50 components, the momentum returns become weaker than those in the other universes excluding the KOSPI 50 constituents. From this observation, the non-KOSPI 50 universes are more effective for implementing more profitable momentum strategies. This effect means that the construction rules of subuniverses play a role of explanatory factors for the momentum returns.
	
	With the subuniverses and universe shrinkage effect, one of the construction rules giving more hints for the source of the momentum returns is the market capitalization which is also an factor of the Fama-French three factor model in analysis of the momentum return. Based on the market capitalization, the size portfolio which buys larger market capitalization equities and sells smaller market-cap equities shows very unique performance which is not reported in any literatures on the momentum effect. It was well-known that a simple negative correlation between the momentum return and market capitalization exists. Instead of the simple negative correlation reported previously, it is found that equities, which have the top 50 to 100th ranks, outperform equities in other size groups. Additionally, this also explains the momentum universe shrinkage effect, why non-KOSPI 50 and $(100-50)$-related market universes beat the pools which include the KOSPI 50 components.
	
	The liquidity also works as a predictive factor on the future returns. The liquidity portfolio ranked by the fractional volume during the estimation period shows the similar results with the traditional momentum strategies in subuniverses. There is also the tendency that performance of the liquidity portfolio becomes weaker when the market universes include the KOSPI 50 members. When the KOSPI 50 companies are not considered as the market universe, the impact of trading is increased whether it is good or bad. This imposes that the interests from investors make the trends to upside or downside directions.
	
	The trading pattern by market participants is also a potential factor to explain the momentum universe shrinkage effect. With three different investor groups such as individual, institutional, and foreign investors, it is found that each subuniverse is governed by the different investor groups. For examples, individual investors prefer the extremes in size during their portfolio selection. Their trades exhibit the strong influences on the large-cap companies such as the KOSPI 50 components and on the small-sized equities in $(200-100)$. It is negatively correlated with future returns. Similar to the individual investors, the institutional investors also like small-cap companies. Additionally, their prediction on the middle-cap companies, which have significant overlap with $(100-50)$, is accurate and the impact is also significant. Meanwhile, the foreigners only focus on the KOSPI 50 companies of which they are good at forecasting the directions. This tendency seems to be related to their limitation on information.
	 
	 In the future, the same test needs to be conducted in the U.S. markets such as S\&P 500 vs. S\&P 100 for a robustness test. Additionally, it will be extended to other stock markets including developed and emerging markets and we will scrutinize any dependence on the momentum universe shrinkage for the momentum strategies. More statistical approaches such as consideration on dummy factors for subuniverses in factor modeling will be also taken in order to pursue more deep understanding on the effect.
	
\section*{Acknowledgement}
	We are thankful to Svetlozar Rachev for useful discussions. We appreciate Jonghyoun Eun for his contribution at the early stage of this work.

\end{document}